\documentclass[11pt,tightenlines,eqsecnum,floats,aps,amsmath,amssymb,nofootinbib,prd,shownopacs,floatfix]{revtex4}

\usepackage{graphicx}
\usepackage{epstopdf}
\usepackage{latexsym}
\usepackage{amssymb}
\usepackage{amsmath}
\usepackage{color}
\usepackage{mathrsfs}
\usepackage{xparse}
\usepackage{float}
\usepackage{mathtools}

\usepackage[center]{subfigure}

\begin{document}

  \renewcommand\arraystretch{2}
 \newcommand{\bq}{\begin{equation}}
 \newcommand{\eq}{\end{equation}}
 \newcommand{\bqn}{\begin{eqnarray}}
 \newcommand{\eqn}{\end{eqnarray}}
 \newcommand{\nb}{\nonumber}
 \newcommand{\lb}{\label}
 \newcommand{\cb}{\color{blue}}
    \newcommand{\cc}{\color{cyan}}
        \newcommand{\cm}{\color{magenta}}
\newcommand{\rc}{\rho^{\scriptscriptstyle{\mathrm{I}}}_c}
\newcommand{\rd}{\rho^{\scriptscriptstyle{\mathrm{II}}}_c} 
\NewDocumentCommand{\evalat}{sO{\big}mm}{%
  \IfBooleanTF{#1}
   {\mleft. #3 \mright|_{#4}}
   {#3#2|_{#4}}%
}
\newcommand{\PRL}{Phys. Rev. Lett.}
\newcommand{\PL}{Phys. Lett.}
\newcommand{\PR}{Phys. Rev.}
\newcommand{\CQG}{Class. Quantum Grav.}
\newcommand{\parallelsum}{\mathbin{\!/\mkern-5mu/\!}}
\newcommand{\blue}{\textcolor{blue}}
\newcommand{\red}{\textcolor{red}}

\title{Primordial scalar power spectrum from the hybrid approach in loop cosmologies}

\author{Bao-Fei Li$^{1}$}
\email{baofeili1@lsu.edu}
\author{Javier Olmedo$^1$ }
\email{jolmedo@phys.lsu.edu}
\author{Parampreet Singh$^1$}
\email{psingh@lsu.edu}
\author{Anzhong Wang$^{2}$}
\email{Anzhong_Wang@baylor.edu}
\affiliation{
$^1$ Department of Physics and Astronomy, Louisiana State University, Baton Rouge, LA 70803, USA\\
$^2$GCAP-CASPER, Department of Physics, Baylor University, Waco, TX, 76798-7316, USA
}

\date{\today}

\begin{abstract}
We compare the primordial scalar power spectra in the loop cosmological models using the effective dynamics of  the hybrid approach to cosmological perturbations in which the background is loop quantized but the perturbations are Fock quantized. The three  loop cosmological models under consideration are the standard LQC, the modified LQC-I (mLQC-I) and the  modified LQC-II (mLQC-II) in the spatially flat Friedmann-Lema\^itre-Robertson-Walker (FLRW) universe with a Starobinsky potential. These models arise from different regularizations of the classical Hamiltonian constraint in the symmetry reduced spacetimes and aim to capture certain features of quantization in loop quantum gravity. When applying the techniques in the hybrid approach to mLQC-I/II, we find the effective Mukhanov-Sasaki equations take the same form as in LQC. The difference among the three models is encoded in the unique expressions of the effective masses in each model. We find that the relative difference in the amplitude of power spectrum between LQC and mLQC-II is approximately $50\%$ in the  infrared and the oscillatory regimes,  whereas this difference can be as large as $100\%$ between mLQC-I and LQC/mLQC-II. Interestingly,  in the  infrared and the oscillatory regimes of mLQC-I, we obtain a suppressed power spectrum from the hybrid approach which is far below the Planck scale. This result is in a striking contrast to the one obtained from the dressed metric approach to perturbations where the corresponding amplitude in this regime is extremely large. Our analysis shows that while the phenomenological predictions are in agreement between the two approaches for LQC and mLQC-II, for mLQC-I the differences between the dressed and hybrid approaches can be quite significant. Our result provides the first robust evidence of difference in predictions between the dressed and hybrid approaches due to respective underlying constructions.

\end{abstract}
\maketitle

\section{Introduction}
\label{Intro}
\renewcommand{\theequation}{1.\arabic{equation}}\setcounter{equation}{0}

The inflationary paradigm not only resolves several long-standing puzzles in the standard big-bang cosmology \cite{ag1981},  but also explains the origin of the  large scale structure in the cosmos \cite{lss}. However, the classical  inflationary spacetimes are past incomplete and   the big bang singularity is inevitable if the solutions are evolved backward to the regime where the energy density of the universe gets close to the Planck scale \cite{BV94}.  In order to extend the inflationary paradigm to the Planckian regime, quantum geometrical effects have to be taken into consideration. One of the successful attempts to achieve this goal is the loop quantum cosmology (LQC) which results from applying the techniques of loop quantum gravity (LQG) to the cosmological settings where a symmetry reduction is first performed before the quantization \cite{review2}. The evolution of the quantum spacetime in LQC  is governed by a non-singular quantum difference equation which results in a resolution of the big bang singularity replacing it with a quantum bounce when spacetime curvature becomes Planckian \cite{aps1,aps2,aps3,acs2010}.  The robustness of this result has been shown for a variety of isotropic and anisotropic spacetimes in the presence of a massless scalar field and constant potentials \cite{review2}. Recently, there is even some progress  on  quantizing the inflationary spacetimes by using the reduced phase space quantization of LQC where the role of the physical clocks is played  by the dust and Klein-Gordon fields  \cite{gls2020} which indicates singularity resolution as in other models of LQC. The phenomenological implications for inflationary background and perturbations in LQC have been studied using the so called effective spacetime description (see \cite{as2017} for a review) whose validity has been verified for isotropic and anisotropic spacetimes \cite{numlsu-1,numlsu-2,numlsu-3,numlsu-4}.

In addition to  the standard LQC in which the Lorentzian term  of the classical  Hamiltonian constraint is treated in the same way as the Euclidean term in the Friedmann-Lema\^itre-Robertson-Walker (FLRW) universe, robustness of the singularity resolution has also been studied with respect to the different quantizations of the classical Hamiltonian constraint in modified LQC. Two notable examples are the so-called  modified LQC-I (mLQC-I) \cite{YDM09,DL17,mehdi,lsw2018} and modified LQC-II (mLQC-II) models \cite{YDM09,lsw2018}. These two models differ from the standard LQC by different regularizations of the Lorentzian term  which result in fourth-order non-singular quantum difference equations \cite{ss6}. It has been shown that in these models, there is a generic resolution of singularity as in LQC \cite{ss5}. The big bang singularity is resolved in the Planck regime and replaced by a quantum bounce and  the inflationary phase can naturally take place with a high likelihood when the inflaton field is coupled to  the gravitational sector with an inflationary potential \cite{lsw2018,lsw2018b,lsw2019}. Although the dynamics in LQC and mLQC-II is qualitatively similar, the difference between LQC and mLQC-I become manifest in the contracting phase  where an emergent quasi-de Sitter space is present in mLQC-I with Planckian values implying that  the contracting phase in mLQC-I is purely a quantum regime without the classical limit. \footnote{Existence of such a phase is not confined to FLRW models but also exists even with standard loop quantization in certain anisotropic spacetimes \cite{djs}.} Given these different regularizations of the Hamiltonian constraint in LQC, an important question is whether the physical predictions resulting from different quantum spacetimes are robust for cosmological perturbations. To answer these questions one needs to carefully understand the way modifications in the Hamiltonian constraint result in modifications to the primordial power spectrum.


 In the literature, there currently exist four primary approaches which address the impacts of the quantum geometry on the primordial power spectra in isotropic LQC (for earlier works see for eg. \cite{pert-old}).  These are the deformed algebra approach \cite{bhks2008,cbgv2012,cmbg2012}, the separate universe approach \cite{wilson2017}, the dressed metric approach \cite{aan2012,aan2013,aan2013b} and the hybrid approach \cite{mm2012, mm2013, gmmo2014,gbm2015,mo2016} (for a recent discussion about similar ideas in anisotropic Bianchi I LQC spacetimes see Refs. \cite{b1-lett,b1-long}).  Among these, the latter two approaches are most widely studied in recent years \cite{d1,d2, bo2016,gbmo2016, tao2017,tao2018,abs2018,nbm2018,nbm2018a}. The dressed metric approach is based on the work by Langlois on the gauge-invariant  perturbations in the  Arnowitt-Deser-Misner (ADM) phase space \cite{lang94} where lapse and shift are treated as Lagrange multipliers.\footnote{This restriction can be lifted in the extended phase space where a generalization of Langlois' treatment has been recently found which allows construction of gauge-invariant variables other than the Mukhanov-Sasaki variable in canonical theory \cite{giesel1,giesel2}.} \footnote{A treatment similar to Langlois' analysis for Bianchi-I spacetimes has been carried out in  \cite{b1-class}.}   In this approach, after expanding the scalar constraint  up to the second order in the perturbations,  the zeroth-order scalar constraint is  loop quantized and the second-order scalar constraint becomes the physical  Hamiltonian that prescribes the dynamics of the inhomogeneous linear perturbations. After quantization, the inhomogeneous gauge invariant degrees of freedom can be interpreted as propagating on a quantum background spacetime which is described by a dressed metric. Furthermore, when the test-field approximation is employed,  in which the background quantum states are sharply peaked around the classical trajectories at late times,  the quantum corrected Mukhanov-Sasaki  equation takes same form as its classical counterpart as long as the relevant background quantities in the equation follow the effective dynamics of  LQC.  Recently, the dressed metric approach has also been extended to  mLQC-I/II with special emphasis on the physical consequences of the regularization ambiguities of the conjugate momentum of the scale factor \cite{lsw2020}. Other relevant work on applying the dressed metric approach to mLQC-I can be found in \cite{IA19}. 
 
 Though the hybrid approach shares a common feature with the dressed metric approach in the sense that perturbations are Fock quantized while the background is loop quantized, it has some important differences.  Based on the work by Halliwell and Hawking \cite{hawking}, in this approach, one usually assumes the spatial geometry to be a three torus and then expands the  spacetime metric and the scalar field on the bases formed by the eigenfunctions of the  Laplace-Beltrami operator compatible with the auxiliary three metric.  After truncating the total Hamiltonian to the second-order in the perturbations, a canonical transformation which concerns both the background variables and the inhomogeneous degrees of freedom is implemented to express the Hamiltonian in terms of the gauge invariant observables, including the Mukhanov-Sasaki  variable, the abelianized linear perturbative constraints and their respective conjugate variables,  while keeping the full canonical structure of the system.  The conjugate momentum of the  Mukhanov-Sasaki  variable is also carefully chosen so that a unitary implementation of the quantum dynamics can be fulfilled \cite{gbm2015, mmo2012}.  Afterwards,  the hybrid quantization ansatz  is employed: the background geometry is loop quantized, the zero-mode of the scalar field is quantized in the standard Schr\"odinger representation while the inhomogeneous perturbations are Fock quantized.  The solution to the resulting quantum dynamical equation is  then solved  by using the Born-Oppenheimer ansatz which approximates the physical state by a direct product of the quantum background state and the states only depending on the gauge invariant modes.  Similar to the dressed metric approach, for the sharply peaked semi-classical background states, there also exists an effective description of the quantum dynamics in the hybrid approach which greatly simplifies the dynamical equations \cite{bo2016}. Recently, the hybrid approach is also applied to
the modified loop cosmological models, such as  mLQC-I \cite{qm2019, gqm2020} for which the time-dependent  mass of the perturbations is analyzed and discussed in \cite{qmp2020}. 
 
 The  goal of this paper is to  study the imprints of the different quantizations of the background  geometry  on the scalar power spectrum in the framework of the hybrid approach. In order to achieve this goal, we apply  the effective dynamics of the hybrid approach in mLQC-I/II to obtain the numerical results of the scalar power spectra from these two models and then compare them with the results from LQC. We assume the gravitational sector of LQC and mLQC-I/II is minimally coupled to  the inflaton field with a Starobinsky inflationary potential whose mass is fixed via the recent Planck 2018 data.  After specifying the initial conditions of the background at the bounce and the initial states of the linear perturbations at some time in the contracting phase, the scalar power spectra are obtained by numerically integrating  the  effective equations of the background and perturbations using the Mathematica internal solver. The results from LQC and mLQC-I/II are then compared from the infrared regime to the ultraviolet  regime of the power spectra. In particular, we find the predictions on the power spectrum of mLQC-I from the hybrid approach is in remarkable contrast  with the results of the same model from the  dressed metric approach in the infrared and oscillatory regimes. Our results show that for LQC and mLQC-II the situation is similar to that in the dressed metric approach, but for mLQC-I there are significant differences in predictions between the two approaches.

This manuscript is organized as follows.  In Sec. \ref{review},  starting from the classical Hamiltonian constraint, we will briefly review the effective dynamics of the hybrid approach in LQC. The Hamilton's equations of the background dynamics and the  Mukhanov-Sasaki equation in LQC  will be given as the basis for the numerical simulations in the following section. In Sec. \ref{mLQC},  we first review the effective dynamics of the background in  mLQC-I/II and then discuss the effective dynamics of the hybrid approach in these two models. We will focus on the  Mukhanov-Sasaki equations from the hybrid approach. In Sec. \ref{power}, based on the results from the previous two sections, we will present the numerical results of the primordial scalar power spectra from the hybrid approach in LQC, mLQC-I and mLQC-II for some representative initial conditions. A comparison among the effective masses and the resulting power spectra from different models and their relative difference will also be given. Finally, in Sec. \ref{summary}, the main results obtained in this paper are summarized.

In our paper, we will use the Planck units with $\hbar=c=1$ while keeping Newton's constant $G$ explicitly. Also, the greek letters are used to denote the 4-dimensional spacetime indices while the Latin letters are for the indices of  the tensors on the 3-dimensional hypersurface. 
 
\section{A brief review  of the hybrid approach in LQC}
\lb{review}
\renewcommand{\theequation}{2.\arabic{equation}}\setcounter{equation}{0}
In this section,  we give a brief review of the hybrid approach in LQC. Since the content has been widely discussed in various articles \cite{nbm2018,bo2016,mo2016,gbmo2016,gbm2015,gmmo2014,mm2013,mop2011}, we only outline the basic ideas and quote the results that are relevant to the purpose of this paper.  In the following,  we will consider a flat FLRW universe with a $\mathbb{T}^3$ topology in which  the four-dimensional globally hyperbolic spacetime is  ADM decomposed into $\mathcal M=\mathcal R\times \mathbb{T}^3$ and the four-metric of the manifold is  parameterized in terms of the lapse $N$, shift $N^i$ and  the three-metric $q_{ij}$ in the ADM decompositions.  Without the inhomogeneities, the homogeneous background in the spatially-flat universe with a $\mathbb{T}^3$ topology is described by 
\bq
\lb{2a1}
\mathrm{d} s^2=-N^2_0(t) \mathrm{d} t^2+a^2(t){}^0h_{ij}\mathrm{d}\theta_i\theta_j,
\eq
where $N_0(t)$ is the lapse function, $a(t)$  the scale factor and ${}^0h_{ij}$ the comoving three-metric. The value of each angular coordinate $\theta_i$ ranges between $0$ and $l_0$ and thus the comoving (physical) volume of the three torus is $l^3_0$ ($a^3 l^3_0$). Any functions defined on the spatial manifold $\mathbb{T}^3$ can be expanded in terms of the eigenfunctions of the Laplace-Beltrami operator compatible with the metric ${}^0h_{ij}$. These eigenfunctions are usually denoted by $\tilde Q_{\vec n, \pm}(\vec \theta)$ with eigenvalues $-\omega^2_n=-4\pi^2\vec n\cdot \vec n/l^2_0$, where $\pm$ stands for the cosine and sine modes respectively, and $\vec n=(n_1,n_2,n_3)\in \nb{Z}^3$ is any tuple of integers with its first component being a strictly positive integer. 

In order to incorporate the inflationary phase driven by a single scalar field, we consider a massive scalar field $\phi$ with the scalar potential $U(\phi)$ minimally coupled to the gravity sector. Following the analysis in \cite{gbm2015},  one can proceed to consider the scalar perturbations around the homogeneous FLRW universe described by the metric (\ref{2a1}).  The inhomogeneities in the lapse, shift and the three-metric can be expanded in the basis of the cosine and sine mode functions $\tilde Q_{\vec n, \pm}$ on the three-torus. The perturbative expansion of the total action of the system which consists of the Einstein-Hilbert action together with the action for a massive scalar field minimally coupled to the gravity is then truncated to the second order in the perturbations, yielding a total Hamiltonian that  is a linear combination of three terms: the first term proportional to the homogeneous mode of the scalar constraint which also includes the quadratic contributions from the linear perturbations, the second term proportional to the perturbed scalar constraint to the first order in perturbations and the third term proportional to the perturbed  momentum constraint to the first order in perturbations. However, this total Hamiltonian is a functional of the gauge variant perturbations, i.e. inhomogeneous degrees of freedom that are not left invariant by the gauge transformation generated by the linear scalar and momentum constraints. 
In order to extract the physical implications from the theory,  it is more convenient to work with the gauge-invariant variables, i.e. the Dirac observables. In general, this can be achieved by a suitable canonical transformation.   In the current context, the appropriate canonical transformations  are introduced in  \cite{gbm2015} in the whole phase space including both homogeneous and inhomogeneous degrees of freedom, which  separate the gauge-invariant variable, namely  the Mukhanov-Sasaki variable denoted in the following by  $\nu_{\vec n, \epsilon}$ and its momentum $\pi_{\nu_{\vec n, \epsilon}}$, from the other  variables  $\nu^{(i)}_{\vec n,\epsilon}$ and their respective momenta $\pi_{\nu^{(i)}_{\vec n,\epsilon}}$ with $i=1,2$. In terms of these new canonical  variables, the total Hamiltonian up to the second order in perturbations can be explicitly written as  \cite{gbm2015}
\bq
\lb{2a2}
\mathcal H_T=\frac{N_0}{16 \pi G} \left(C_0+ \sum_{\vec n, \epsilon}C^{\vec n, \epsilon}_2\right)+ \sum_{\vec n, \epsilon} G_{\vec n, \epsilon} \pi_{\nu^{(1)}_{\vec n,\epsilon}}+ \sum_{\vec n, \epsilon} K_{\vec n, \epsilon} \pi_{\nu^{(2)}_{\vec n,\epsilon}},
\eq
where $G_{\vec n, \epsilon}$ and $K_{\vec n, \epsilon} $  are the the coefficients of the Fourier modes of the linear perturbations of the lapse and shift. Besides, $\pi_{\nu^{(1)}_{\vec n,\epsilon}}$ and $\pi_{\nu^{(2)}_{\vec n,\epsilon}}$ are equivalent to the perturbed scalar and momentum constraints which are linear in perturbations. When the theory is quantized by following the  Dirac quantization approach, the physical states will be independent of  $\nu^{(1)}_{\vec n,\epsilon}$  and $\nu^{(2)}_{\vec n,\epsilon}$. As a result, the sector $(\nu^{(i)}_{\vec n,\epsilon}, \pi_{\nu^{(i)}_{\vec n,\epsilon}})$ is decoupled from the physical one.
The first term in the total Hamiltonian only concerns the homogeneous background and the  Mukhanov-Sasaki variable and its momentum, which is explicitly given by
\bq 
\lb{2a3}
\mathcal H_\mathrm{MS}=\frac{N_0}{16 \pi G} \left(C_0+ \sum_{\vec n, \epsilon}C^{\vec n, \epsilon}_2\right),
\eq
where the subscript `$\mathrm{MS}$' implies that the Hamiltonian  $\mathcal H_\mathrm{MS}$ generates the dynamics of the  Mukhanov-Sasaki variable and its momentum.  The unperturbed scalar constraint is given by 
\bq
\lb{2a4}
C_0=-\frac{6}{\gamma^2 }\frac{\Omega^2}{v}+8\pi G\left(\frac{p^2_\phi}{v}+2 v U(\phi)\right),
\eq
where $\gamma$ is the Barbero-Immirzi parameter  which is usually set to $0.2375$ from the black hole thermodynamics in LQG, $p_\phi$ is the conjugate momentum of the scalar field and $U(\phi)$ represents the potential of the scalar field. For the geometrical degrees of freedom, instead of the scale factor and its momentum, we use the variables $(v, b)$ which will be more convenient  for our later discussion of the effective dynamics   in the loop cosmological models.  In the classical theory, $\Omega=vb$ with $v$ representing the physical volume of the 3-torus and $b=\gamma H$ where $H$ is the Hubble parameter. Meanwhile,  $C^{\vec n, \epsilon}_2$ denotes the quadratic  corrections from the modes labeled by $(\vec n , \epsilon)$, which takes the form  \cite{gmmo2014, bo2016}
\bq
\lb{2a5}
C^{\vec n, \epsilon}_2=\frac{8\pi G}{v^{1/3}}\left(\pi^2_{\nu_{\vec n, \epsilon}}+E^n \nu^2_{\vec n, \epsilon}\right),
\eq
with 
\bqn
\lb{2a6}
E^n&=&\omega^2_n+s,\\
\lb{2a7}
s&=&\frac{4 \pi G p^2_\phi}{3 v^{4/3}}\left(19-24 \pi G \gamma^2 \frac{p^2_\phi}{\Omega^2}\right)+ v^{2/3}\left(U_{, \phi\phi}+\frac{16 \pi G \gamma p_\phi }{\Omega}U_{,\phi}-\frac{16\pi G}{3}U\right),
\eqn
where $U_{,\phi}\equiv \partial U/\partial \phi$ and so on.
With the  Poisson brackets given by
\bq
\lb{2a8}
\{b,v\}=4\pi G \gamma, \quad \quad \{\phi, p_\phi\}=1, \quad \quad \{\nu_{\vec n, \epsilon},\pi_{\nu_{\vec n, \epsilon}} \}=\delta_{\vec n \vec n' }\delta_{\epsilon \epsilon'},
\eq
it is straightforward to find the Hamilton equations for the canonical variables and  their respective momenta. However, different quantizations of the geometric sector in LQC can result in different forms of $\Omega$ in the effective description of the quantum  dynamics. In order to cast the Hamilton equations into the most general form which will also be  valid in the modified LQC models, we will keep $\Omega$, as a function of $v$ and $b$, explicit in the equations. Hence, when  ignoring the back-reaction of the perturbations on the homogeneous and isotropic background,  the evolution of the background dynamics obeys the following equations 
\bqn
\lb{h1}
\dot v&=&N_0\frac{3\Omega}{v\gamma}\frac{\partial \Omega}{\partial b},\\
\lb{h2}
\dot b&=&\frac{3N_0\Omega^2}{2v^2\gamma}-\frac{3N_0\Omega}{\gamma v}\frac{\partial \Omega}{\partial v}-4\pi G\gamma N_0 P,\\
\lb{matter1}
\dot \phi &=&N_0\frac{p_\phi}{v}, \\
\lb{matter2}
\quad \dot p_\phi&=&-N_0v U_{,\phi},
\eqn
where $P$ denotes the pressure of the scalar field which is given by 
\bq
P=\frac{p^2_\phi}{2v^2}-U.
\eq  
Meanwhile, the time evolution of the scalar modes $\nu_{\vec n, \epsilon}$ is governed by 
\bq
\lb{numeric}
{\dot \nu}_{\vec n, \epsilon}=\frac{N_0}{v^{1/3}}\pi_{\nu_{\vec n, \epsilon}}, \quad \quad {\dot \pi}_{\nu_{\vec n, \epsilon}}=-\frac{N_0E^n \nu_{\vec n, \epsilon}}{v^{1/3}}.
\eq
In the above formulae, if the lapse $N_0$ is taken to be $v^{1/3}$, then the overdots represent the differentiations with respect to the conformal time, in which the equation of motion of each scalar mode takes the form
\bq
\lb{2a9}
\nu^{\prime\prime}_{\vec n, \epsilon}+\left(\omega^2_n+s\right)\nu_{\vec n, \epsilon}=0,
\eq
where a prime denotes the differentiation with respect to the conformal time and $s$ is given in (\ref{2a7}).
In terms of the Mukhanov-Sasaki  variable  $Q_k=\nu_k/a$, the above equation  is equivalent to
\bq
\lb{2a10}
\ddot Q_k+3H\dot Q_k+\left(\frac{\omega^2_n+s}{a^2}+H^2+\frac{\ddot a }{a}\right)Q_k=0,
\eq
where $H$ is the Hubble rate, as mentioned above, and the derivatives of the relevant quantities are with respect to the cosmic time when the lapse $N_0$ is set to unity.    The above equations (\ref{2a10})  and (\ref{h1})-(\ref{matter2}) constitute a fundamental set of the equations which describe the dynamics of both background and linear perturbations in the hybrid approach at the classical level.  For the pragmatic purpose,  in the following, the discrete spectra $\omega^2_n$ is set equal to the continuous comoving wavenumber $k$  which is equivalent to taking the limit $l_0\rightarrow \infty$.

\subsection{The hybrid quantization}
In the hybrid approach, the quantization of the Hamiltonian constraint (\ref{2a3}) is implemented in two successive steps which involves certain assumptions. First, the homogeneous gravitational sector is loop quantized in the $\bar \mu$ scheme in LQC and the matter sector is quantized in the usual Schr\"odinger  representation. Note that in the Dirac quantization in LQC, the quantization of background is not yet available in the presence of a potential. As a result one generally assumes, as in the dressed metric approach, an existence of background quantization with a physical inner product generally taken to be the same as in absence of potentials. In the following we work with same assumption as being made in previous works but note that this limitation can be overcome given recent developments to include a potential in the reduced phase space quantization \cite{gls2020}. Second, as in the dressed metric approach, the inhomogeneous degrees of freedom are not loop but Fock quantized. As a result, the kinematic Hilbert space is a tensor product of the individual Hilbert space  for each sector, that is, $\mathcal H_\mathrm{kin}=\mathcal H^\mathrm{grav}_\mathrm{kin}\otimes \mathcal H^\mathrm{matt}_\mathrm{kin}\otimes \mathcal F$. More specifically, the kinematic Hilbert space of the homogeneous gravitational sector  is  $\mathcal H^\mathrm{grav}_\mathrm{kin}=L^2(\mathbb{R}_\mathrm{Bohr},d \mu_\mathrm{Bohr})$ where $\mathbb{R}_\mathrm{Bohr}$ is the Bohr compactification of the real line and $d \mu_\mathrm{Bohr}$ its Haar measure.  $ \mathcal H^\mathrm{grav}_\mathrm{kin}$ is spanned by the eigenstates of the volume operator which are usually denoted by $\{|v\rangle, v\in \mathbb{R} \}$ with the discrete norm $\langle v_1| v_2\rangle=\delta_{v_1,v_2}$. The fundamental operators in $\mathcal H^\mathrm{grav}_\mathrm{kin}$ in the $\bar \mu$ scheme in LQC \cite{aps3} are the volume operator $\hat v$ and the holonomy operator $\hat N_{\bar \mu}=\widehat{e^{-i \lambda b/2}}$ with $\lambda =\sqrt{\Delta}$ and $\Delta$($=4\sqrt{3}\pi\gamma l^2_\mathrm{Pl}$) is the minimum area eigenvalue in LQG. 

In the hybrid approach, one usually considers the Martin-Benito-Mena Marugan-Olmedo prescription \cite{bmo2009} for the factor ordering in the Hamiltonian  constraint operator. With this prescription, the zero mode of the homogeneous sector is represented by 
\bq
\lb{2a11}
\hat C_0=\left(\widehat{\frac{1}{v}}\right)^{1/2}\left(-\frac{6}{\gamma^2} \hat \Omega^2 +8\pi G \hat p^2_\phi+2 \hat v^2 U(\hat \phi)\right)\left(\widehat{\frac{1}{v}}\right)^{1/2},
\eq 
here $\hat \phi$ and $\hat p_\phi$ ($=-i\hbar \frac{\partial}{\partial \phi}$) are the operators in the kinematic Hilbert space of the matter sector which is $\mathcal H^\mathrm{matt}_\mathrm{kin}=L^2(\mathbb{R}, d\phi)$. The operator $\hat \Omega$ is given by 
\bq
\lb{2a12}
\hat \Omega=\frac{1}{4 i \sqrt{\Delta}}\hat v^{1/2}\left(\widehat{\mathrm{sgn}(v)}\left(\hat N_{2\bar \mu}-\hat N_{-2\bar \mu}\right)+\left(\hat N_{2\bar \mu}-\hat N_{-2\bar \mu}\right)\widehat{\mathrm{sgn}(v)}\right)\hat v^{1/2}.
\eq
The operator $\hat \Omega^2$ annihilates the zero volume state $|v=0\rangle$ and selects a group of separable subspaces $\mathcal H^\pm_\epsilon$ which are formed by the states with support on the lattices ${\cal L}_{\pm \epsilon} = \{ v = \pm (4n + \epsilon)\}$ with $n \in \mathbb{N}$ and $\epsilon \in (0,4]$. The action of $\hat \Omega^2$, as well as $\hat C_0$, leaves these subspaces invariant and do not mix  states with support on the  opposite signs of the volume.  

On the other hand, the inhomogeneous sector  is quantized in the Fock representation by choosing the annihilation-like variable 
\bq
a_{v_{\vec n, \epsilon}}=\frac{1}{\sqrt{2\omega_n}}\left(\omega_n v_{\vec n, \epsilon} +i \pi_{v_{\vec n ,\epsilon}}\right),
\eq
and its complex conjugate $a^*_{v_{\vec n, \epsilon}}$ as the creation-like variable. The quantization is then implemented by promoting these variables to the annihilation and creation operators. The resulting Fock space is spanned by the direct products of the eigenstates of the occupation number operator $\mathcal N_{\vec n, \epsilon}$ for each mode $(\vec n ,\epsilon)$. Finally, the physical states described by $\Psi(v, \phi, \mathcal N)$ should be annihilated by the quantum Hamiltonian constraint 
\bq
\hat {\mathcal H}_\mathrm{MS}=\frac{1}{16 \pi G} \left(\hat C_0+ \sum_{\vec n, \epsilon}\hat {C}^{\vec n, \epsilon}_2\right),
\eq
where we take $N_0=1$ at the classical level and $\hat C_0$ is given in (\ref{2a11}). Here the operator $\hat {C}^{\vec n, \epsilon}_2$ is promoted from its classical counterpart (\ref{2a5}) whose  explicit form can be  found in \cite{gbm2015}. Here, we want to emphasize that to obtain  $\hat {C}^{\vec n, \epsilon}_2$, the $\Omega^2$ term in the effective mass  (\ref{2a7}) is promoted to  $\hat \Omega^2$ given by the square of (\ref{2a12}). However, the $1/\Omega$ term in the effective mass can not be directly promoted to the desired operator as  $ \hat \Omega$ is a difference operator which only translates eigenstates $| v\rangle$ by two units. In order to make  $1/\Omega$  not mix the states from different  superselection  subspaces, the following prescription is used in hybrid approach:
\bq
\frac{1}{\Omega}\rightarrow \hat \Omega^{-1} \hat \Lambda  \hat  \Omega^{-1},
\eq
with $\hat \Lambda $ given by 
\bq
\lb{2a13}
\hat \Lambda=\frac{1}{8 i \sqrt{\Delta}}\hat v^{1/2}\left(\widehat{\mathrm{sgn}(v)}\left(\hat N_{4\bar \mu}-\hat N_{-4\bar \mu}\right)+\left(\hat N_{4\bar \mu}-\hat N_{-4\bar \mu}\right)\widehat{\mathrm{sgn}(v)}\right)\hat v^{1/2}.
\eq
As compared with $\hat \Omega$, $\hat \Lambda$ is defined with holonomies of double fiducial length and hence  preserves the superselection sectors.  For the other homogeneous factors in the effective mass, a symmetric factor ordering is employed. Finally, the physical quantum states is governed by 
\bq
\hat {\mathcal H}_\mathrm{MS}  \Psi(v, \phi, \mathcal N)=0.
\eq
In general, it is very difficult to solve this equation for physical quantum states  (see \cite{mm2013} for a specific algorithm though). Approximated solutions where one adopts a Born-Oppenheimer ansatz have been studied \cite{gmmo2014,gbm2015,mo2016}. Here, under some reasonable approximations, one can derive a dressed metric formulation for both scalar and tensor perturbations. But, in practice, one usually focuses on sharply peaked states and hence turns into the effective description of the quantum dynamics to extract the physical implications of the theory.

\subsection{Effective dynamics in the hybrid approach }
In LQC, although the Schr\"odinger equation  for the physical quantum states is a non-singular quantum difference equation, the effective description of the quantum spacetime for the semi-classical states which are sharply peaked around the classical solutions at late times have proved to accurately  capture the properties of the quantum evolution in LQC for a variety of isotropic and anisotropic models \cite{VT08, aps2, numlsu-2, numlsu-3, numlsu-4}. For the spatially-flat model,  it turns out that the effective description of the background dynamics  is based on an effective Hamiltonian which can be obtained by replacing the momentum variable $b$ with $\sin(\lambda b)/\lambda$  in the classical Hamiltonian (\ref{2a3}).  This substitution can be obtained from the operator (\ref{2a12})  for the semi-classical states in which the expectation values of products of operators are replaced with the products of expectation values of the same operators.  As a result, in LQC, the equations of motion for the effective background dynamics are given in (\ref{h1})-(\ref{matter2}) with $\Omega$ given by 
\bq
\Omega_\mathrm{LQC}=v\frac{\sin(\lambda b)}{\lambda}.
\eq
In the classical limit, $\lambda b\ll1$ which reduces $\Omega_\mathrm{LQC}$ to its classical expression $\Omega=vb$. Therefore, the equations of motion for the background dynamics in LQC take the form
\bqn
\lb{lqc1}
\dot v&=&\frac{3v}{2\lambda \gamma}\sin(2\lambda b), \\
\lb{lqc2}
\dot b&=&-\frac{3\sin^2\left(\lambda b\right)}{2 \gamma \lambda^2}-4\pi G\gamma P,
\eqn
where $N_0$ is set to unity and overdots represent differentiation with respect to the cosmic time. Note that the equations of motion in matter sector are still given by (\ref{matter1})-(\ref{matter2}).  

Similarly, the effective dynamics of the scalar perturbations is prescribed by the Mukhanov-Sasaki equations (\ref{2a9}) and (\ref{2a10}) under the conditions that: (i)  the evolution of all the relevant  background quantities agrees with their effective dynamics  described in (\ref{matter1})-(\ref{matter2}) and  (\ref{lqc1})-(\ref{lqc2}); (ii) as in the quantum theory, in the effective mass $s$ given in (\ref{2a7}), the $1/\Omega^2$ and $1/\Omega$ is given by their effective expressions  which in the semi-classical limit take the form \cite{bo2016}
\bq
\lb{2b1}
\frac{1}{\Omega^2}\rightarrow \frac{1}{\Omega^2_\mathrm{LQC}}, \quad \quad \frac{1}{\Omega}=\frac{\Lambda_\mathrm{LQC}}{\Omega^2_\mathrm{LQC}},
\eq
with 
\bq
\lb{2b2}
\Lambda_\mathrm{LQC}=v\frac{\sin(2\lambda b )}{2\lambda}.
\eq
Here $\Lambda_\mathrm{LQC}$ is the semi-classical limit of the operator (\ref{2a13}) for the highly peaked semi-classical states. As a result, the effective mass in the Mukhanov-Sasaki equation in LQC is explicitly given by 
\bq
s=\frac{4 \pi G p^2_\phi}{3 v^{4/3}}\left(19-24 \pi G \gamma^2 \frac{p^2_\phi}{\Omega^2_\mathrm{LQC}}\right)+ v^{2/3}\left(U_{, \phi\phi}+\frac{16 \pi G \gamma p_\phi \Lambda_\mathrm{LQC}}{\Omega^2_\mathrm{LQC}}U_{,\phi}-\frac{16\pi G}{3}U\right).
\eq
As now we are equipped with a complete set of dynamical equations for both background and scalar perturbations, the scalar power spectrum can be obtained through numerical simulations as long as the initial conditions and initial states are specified. 

To summarize, we have discussed the basic ideas in the hybrid approach in LQC and gave the fundamental Hamilton equations for the background and the Mukhanov-Sasaki equation of the scalar perturbations in the effective description of the quantum theory. This effective dynamics is based on the Born-Oppenheimer ansatz and the assumption that there exist  some semi-classical states in which the effective equations of motion of the expectation values of the fundamental observables are consistent with the effective dynamics in LQC. We will employ the same ansatz and the assumption in the next section to obtain the effective equations of motion for both of the background and the scalar perturbations in the modified LQC models. 

\section{The modified loop quantum cosmology and the hybrid approach}
\lb{mLQC}
\renewcommand{\theequation}{3.\arabic{equation}}\setcounter{equation}{0}
In this section, we briefly review two modified LQC models, namely mLQC-I and mLQC-II. We will focus on their effective dynamics and give their respective  Hamilton's equations for the background evolution and the  relevant equations for the scalar perturbations in the hybrid approach. We follow the conventions in Refs. \cite{lsw2018,lsw2018b,lsw2019} which can be referred to for further discussion.
\subsection{mLQC-I}
The mLQC-I model was first proposed as an alternative quantization of the Hamiltonian constraint in a spatially-flat FLRW universe \cite{YDM09}. It was later rediscovered in \cite{DL17} by computing the expectation values of the Hamiltonian constraint with the complexifier coherent states. Phenomenologically, this model is characterized by an asymmetric bounce with its contracting phase quickly tending to a  quasi de Sitter phase with an effective Planck-scale cosmological constant \cite{mehdi} and a rescaled Newton's constant \cite{lsw2018}. Similar to the standard LQC, we assume the validity of the effective description of the quantum spacetime. The effective dynamics in mLQC-I can be obtained from an effective Hamiltonian which can be arrived at by the prescription
\bq
\lb{3a1}
\Omega^2_{{\scriptscriptstyle{\mathrm{I}}}}=-\frac{v^2\gamma^2}{\lambda^2}\Big\{\sin^2\left(\lambda b\right)-\frac{\gamma^2+1}{4\gamma^2} \sin^2\left(2 \lambda b\right) \Big\}.
\eq
Substituting the above expression of $\Omega^2_{{\scriptscriptstyle{\mathrm{I}}}}$ into (\ref{h1}) and (\ref{h2}), one finds the Hamilton's equations for the effective dynamics in mLQC-I:
\bqn
\lb{mLQCIa}
\dot v&=&\frac{3v\sin(2\lambda b)}{2\gamma \lambda}\Big\{(\gamma^2+1)\cos(2\lambda b)-\gamma^2\Big\},\\
\lb{mLQCIb}
\dot b&=&\frac{3\sin^2(\lambda b)}{2\gamma \lambda^2}\Big\{\gamma^2\sin^2(\lambda b)-\cos^2(\lambda b)\Big\} -4\pi G\gamma  P,
\eqn
where $N_0$ is set to unity and the overdots in the evolution equations represent the differentiation with respect to the cosmic time. Besides, the equations of motion of the matter sector still take the form of (\ref{matter1}) and (\ref{matter2}) as long as the lapse in those equations is set to unity. From the Hamilton equations, it is straightforward to derive the Friedmann equation in mLQC-I. The Friedmann equation develops two distinctive expressions in the contracting and the expanding phases resulting in an asymmetric bounce  (for their exact forms and details see \cite{lsw2018}). In mLQC-I, the bounce takes place when the energy density reaches its maximum value at 
\bq
\lb{3a2}
\rho=\rho_c^{{\scriptscriptstyle{\mathrm{I}}}} \equiv \frac{\rho_c}{4\left(\gamma^2+1\right)}. 
\eq
Similar to LQC, the momentum $b$ in mLQC-I is also a monotonically decreasing function in the forward evolution which ranges between $\Big[0, \frac{1}{\lambda}\arcsin(\sqrt{1/(\gamma^2+1)})\Big]$ and equals $\frac{1}{\lambda}\arcsin(\sqrt{1/(2\gamma^2+2)})$ at the bounce.  

For mLQC-I the hybrid approach for the primordial power spectrum  has been studied earlier in \cite{qm2019,gqm2020}. Similar to LQC, the kinematic Hilbert space is a direct product of the three subspaces, namely, $\mathcal H_\mathrm{kin}=\mathcal H^\mathrm{grav}_\mathrm{kin}\otimes \mathcal H^\mathrm{matt}_\mathrm{kin}\otimes \mathcal F$. However, in mLQC-I, the gravitational sector of the quantum  Hamiltonian constraint changes. In particular, the operator $\hat \Omega^2 $ now becomes \cite{qm2019,gqm2020}
\bq
\lb{3a3}
\hat {\Omega}^2_{{\scriptscriptstyle{\mathrm{I}}}}=-\gamma^2\left(\hat \Omega^2_{2\bar \mu}-\frac{\gamma^2+1}{4\gamma^2}\hat \Omega^2_{4\bar \mu}\right),
\eq
where the subscript of $\hat {\Omega}^2_{{\scriptscriptstyle{\mathrm{I}}}}$ indicates it is the   $\hat \Omega^2 $ operator in mLQC-I. In the above formula, we defined for an arbitrary integer $n$
\bq
\lb{3a4}
\hat \Omega_{n\bar \mu}=\frac{1}{4 i \sqrt{\Delta}}\hat v^{1/2}\left(\widehat{\mathrm{sgn}(v)}\left(\hat N_{n\bar \mu}-\hat N_{-n\bar \mu}\right)+\left(\hat N_{n\bar \mu}-\hat N_{-n\bar \mu}\right)\widehat{\mathrm{sgn}(v)}\right)\hat v^{1/2},
\eq
with $\hat N_{n\bar \mu}=\widehat {e^{-i n \lambda b/2}}$. The operator $\hat {\Omega}^2_{{\scriptscriptstyle{\mathrm{I}}}}$ is also compatible with the same  superselection subspace $\mathcal H^\pm_\epsilon$ with support on the lattices with step four. As a result, the operator $\hat \Lambda$ can be chosen in the same form as in LQC which turns out to be
\bq
\lb{3a5}
\hat \Lambda_{{\scriptscriptstyle{\mathrm{I}}}}= \hat \Omega_{4\bar \mu}/2.
\eq
In the effective description of quantum dynamics, the evolution of the scalar perturbations in mLQC-I is  prescribed by the same form of the Mukhanov-Sasaki  equation in (\ref{2a9}) and (\ref{2a10}) under the  following conditions: (i) the evolution of the homogeneous background quantities are now governed by the effective equations (\ref{matter1})-(\ref{matter2}) and (\ref{mLQCIa})-(\ref{mLQCIb}); (ii)  in the effective mass $s$, the following substitutions are employed
\bq
\frac{1}{\Omega^2}\rightarrow \frac{1}{\Omega^2_{{\scriptscriptstyle{\mathrm{I}}}}}, \quad \quad  \frac{1}{\Omega}\rightarrow \frac{\Lambda_{{\scriptscriptstyle{\mathrm{I}}}}}{\Omega^2_{{\scriptscriptstyle{\mathrm{I}}}}},
\eq
where  $\Lambda_{{\scriptscriptstyle{\mathrm{I}}}}$ is the expectation value of the operator $\hat \Lambda_{{\scriptscriptstyle{\mathrm{I}}}}$ for the sharply peaked  semiclassical states. It takes the same form as $\Lambda_\mathrm{LQC}$ given by (\ref{2b2}). As a result, the effective mass of the  Mukhanov-Sasaki  equation  (\ref{2a9}) and  (\ref{2a10}) in mLQC-I is explicitly given by
\bq
s=\frac{4 \pi G p^2_\phi}{3 v^{4/3}}\left(19-24 \pi G \gamma^2 \frac{p^2_\phi}{\Omega^2_{{\scriptscriptstyle{\mathrm{I}}}}}\right)+ v^{2/3}\left(U_{, \phi\phi}+\frac{16 \pi G \gamma p_\phi \Lambda_{{\scriptscriptstyle{\mathrm{I}}}}}{\Omega^2_{{\scriptscriptstyle{\mathrm{I}}}}}U_{,\phi}-\frac{16\pi G}{3}U\right),
\eq
with  $\Omega^2_{{\scriptscriptstyle{\mathrm{I}}}}$  given in  (\ref{3a1}). 

\subsection{mLQC-II}
The mLQC-II model  was also first proposed in \cite{YDM09} as a different quantization of the classical Hamiltonian in the spatially-flat FLRW universe. Its effective dynamics and implications on the inflationary paradigm were later studied in detail  in \cite{lsw2019, lsw2018b}.  Similar to the standard LQC, the evolution of the universe in mLQC-II  is symmetric with respect to the bounce when only a massless scalar field is coupled to the gravitational sector. Its effective dynamics can be described by an effective Hamiltonian constraint which leads to the Hamilton equations in the same form as (\ref{h1}) and (\ref{h2}) as long as $\Omega^2$ is replaced by its corresponding form in mLQC-II given by 
\bq
\lb{3a3}
\Omega^2_{{\scriptscriptstyle{\mathrm{II}}}}=\frac{4v^2}{\lambda^2}\sin^2\left(\frac{\lambda b}{2}\right)\Big\{1+\gamma^2 \sin^2\left(\frac{\lambda b}{2}\right) \Big\}.
\eq
Correspondingly, the Hamilton equations in mLQC-II read
\bqn
\lb{mLQCIIa}
\dot v&=&\frac{3v\sin(\lambda b)}{\gamma \lambda}\Big\{1+\gamma^2-\gamma^2\cos\left(\lambda b\right)\Big\}, \\
\lb{mLQCIIb}
\dot b&=&-\frac{6\sin^2\left(\frac{\lambda b}{2}\right)}{\gamma \lambda^2}\Big\{1+\gamma^2\sin^2\left(\frac{\lambda b}{2}\right)\Big\}-4\pi G\gamma P,
\eqn
where the lapse is set to unity and the overdots represent the differentiation with respect to the cosmic time. 
In mLQC-II, the bounce takes place when the energy density reaches its maximum value at 
\bq
\rho=\rho_c^{{\scriptscriptstyle{\mathrm{II}}}} \equiv 4(\gamma^2+1)\rho_c.
\eq
The momentum $b$ in mLQC-II monotonically decreases in the forward evolution of the universe from $2\pi/\lambda$ to $0$ and equals $\pi/\lambda$ at the bounce.  The quantization of the homogeneous and inhomogeneous sectors can be carried out in a similar way as in LQC, and the only difference lies in the gravitational sector which due to a difference in quantization corresponds to a different operator in the kinematic Hilbert space. More specifically, the $\hat \Omega^2$ operator in mLQC-II takes the form
\bq
\hat \Omega^2_{{\scriptscriptstyle{\mathrm{II}}}}=4 \hat \Omega^2_{\bar \mu}+4\gamma^2\lambda^2\left(\widehat{\frac{1}{v}} \right)\hat \Omega^4_{\bar \mu}\left(\widehat{\frac{1}{v}} \right),
\eq
which is compatible with the superselection subspaces with support on the semilattices with step two. As a result, in order for the total Hamiltonian constraint $\hat{ \mathcal H}_\mathrm{MS}$ to be well-defined in the same subspaces, the $\hat \Lambda$ operator can be chosen as a quantum difference operator that  translates the eigenstates $| v\rangle$ with a displacement of any multiples of two. The simplest choice in this case would be 
\bq
\hat \Lambda_{{\scriptscriptstyle{\mathrm{II}}}}=\hat \Omega_{2\bar \mu},
\eq
which in the effective dynamics corresponds to 
\bq
\Lambda_{{\scriptscriptstyle{\mathrm{II}}}}=v \frac{\sin\left(\lambda b\right)}{\lambda}.
\eq
Together with $\Omega^2_{{\scriptscriptstyle{\mathrm{II}}}}$ in (\ref{3a3}),  it determines the exact form of  the effective mass in the Mukhanov-Sasaki equation (\ref{2a10}) in mLQC-II, which reads
\bq
s=\frac{4 \pi G p^2_\phi}{3 v^{4/3}}\left(19-24 \pi G \gamma^2 \frac{p^2_\phi}{\Omega^2_{{\scriptscriptstyle{\mathrm{II}}}}}\right)+ v^{2/3}\left(U_{, \phi\phi}+\frac{16 \pi G \gamma p_\phi \Lambda_{{\scriptscriptstyle{\mathrm{II}}}}}{\Omega^2_{{\scriptscriptstyle{\mathrm{II}}}}}U_{,\phi}-\frac{16\pi G}{3}U\right).
\eq

Thus, focusing on the effective description of the  hybrid approach in LQC, both the Hamilton equations of the background dynamics and the Mukhanov-Sasaki  equations of the scalar perturbations can be obtained in mLQC-I and mLQC-II. In particular, we have specified the exact forms of the operators $\hat \Omega$ and $\hat \Lambda$, as well as their counterparts in the effective dynamics in each model. These equations will be used in the numerical simulations of the primordial scalar power spectrum in mLQC-I/II in the next section.

\section{Primordial power spectrum from the hybrid approach in modified loop quantum cosmology}
\lb{power}
\renewcommand{\theequation}{4.\arabic{equation}}\setcounter{equation}{0}
In this section, based on the effective dynamics of the background and the perturbations introduced in the previous sections, we  proceed with the numerical simulations and compare the scalar power spectra between LQC and mLQC-I/II in the hybrid approach. Moreover, we will also compare the difference between the dressed metric approach and the hybrid approach in the context of mLQC-I where the de Sitter  phase differentiates these two approaches by allowing for different types of the initial conditions in the contracting phase. Here we would use the results in our previous paper in the dressed metric approach \cite{lsw2020}.

We will start with the fixation of the free parameter in the inflationary model. Based on the Planck 2018 data which  favors an inflationary potential with a plateau,  we choose the scalar potential $U(\phi)$ to be the Starobinsky potential  which is explicitly given by 
\bq
\lb{4.1}
U=\frac{3m^2}{32\pi G}\left(1-e^{-\sqrt{\frac{16\pi G}{3}} \phi}\right)^2.
\eq
Due to the almost flat right wing of the potential, the tensor-to-scalar ratio predicted in the inflationary models with the Starobinsky potential fits the observational data very well. The pivot mode is chosen at $k^0_*=0.05\,({\rm Mpc})^{-1}$ where the superscript `0' refers to the value at present. With  the scalar power spectrum $A_s$ and the scalar spectral index $n_s$ given respectively by \cite{Planck2018}
\bq
\lb{4.2}
\ln (10^{10}A_s)=3.044\pm0.014  ~(68\% \mathrm{CL}) ,\quad\quad n_s=0.9649\pm0.0042 ~(68\% \mathrm{CL}),
\eq
one can fix the mass of the scalar field to be $m=2.44\times10^{-6}$. Some of the relevant observables at the horizon crossing during inflation can also be computed, which are 
\bq
\lb{4.3}
\phi_*=1.07,\quad \quad \dot \phi_*=-5.02\times10^{-9},\quad \quad H_*=1.20\times10^{-6}.
\eq
Since the Hubble rate decreases during the slow-roll inflation, the moment for the horizon exit of the pivot mode denoted by $t_*$ is then determined when the Hubble rate decreases to the value of $H_*$ in the slow-roll  phase of our numerical solutions. 

In addition, all our simulations were performed using a combination of the StiffnessSwitching and ExplicitRungeKutta numerical  methods in Mathematica. The background solutions were obtained  from numerical integrations of the modified Friedmann equations in LQC and mLQC-I/II, while  the primordial power spectrum for the linear perturbations was found from numerically integrating (\ref{numeric}) with the respective effective mass in each model.

\subsection{The initial conditions of the background and the initial states of the scalar perturbations}

In our simulations, the initial conditions of the background dynamics are chosen at the bounce where the energy density reaches its maximum value. Due to the rescaling freedom in volume, we choose 
$v_0=1$ for our numerical solutions. The canonical variable $b$ is fixed at the bounce. More specifically, as  discussed in  the last section, at the bounce,  $b_0=\frac{\pi}{2\lambda}$ in LQC;   $b_0=\mathrm{arcsin}(\sqrt{1/(2\gamma^2+2)})/\lambda$ in mLQC-I; $b_0=\frac{\pi}{\lambda}$ in mLQC-II. The degrees of freedom in the matter sector consists of $\phi$ and $p_\phi$. When the energy density reaches its maximum value,
\bq
\lb{4a1}
\rho=\frac{p^2_{\phi_0}}{2v_0^2}+U(\phi_0)=\rho^i_c,
\eq
here the subscript `$0$' indicates the values of the relevant quantities are set at the bounce and $\rho^i_c$ stands for the maximum energy density in LQC and mLQC-I/II. 

With regard to the initial states of the scalar perturbations, they are chosen at some finite time in the contracting phase. In general,  the choice of the initial states  is based on their equation of motion 
\bq
\lb{4a2}
\nu_k^{\prime \prime}+\left(k^2+s\right)\nu_k=0,
\eq
where $s$ is the effective mass and the mode function satisfies the Wronskian condition
\bq
\lb{wronskian}
\nu_k(\nu^\prime_k)^*-(\nu_k)^*\nu^\prime_k=i,
\eq
with the asterisk standing for the complex conjugate.
 As discussed in \cite{lsw2020}, the initial states in the contracting phase can be chosen as the adiabatic states, given explicitly by the WKB solutions of (\ref{4a2}),
\bq
\lb{4a3}
\nu_k=\frac{1}{\sqrt{2 W_k}}e^{-i \int^\eta W_k(\bar \eta)d\bar \eta}.
\eq
Substituting the above solution back into (\ref{4a2}), one can find an iterative equation for $W_k$. Then, starting from the zeroth order  solution, $W^{(0)}_k=k$,  the adiabatic solutions at the second and fourth orders can be obtained as
\bq
\lb{4a4}
W^{(2)}_k=\sqrt{k^2+s}, \quad \quad  W^{(4)}_k=\frac{\sqrt{f(s,k)}}{4|k^2+s|}.
\eq
Here $f(s,k)=5s'^2+16k^4(k^2+3s)+16s^2(3k^2+s)-4s^{''}(s+k^2)$.
For any two sets of the initial states, say $\{\nu_k\}$ and $\{\mu_k\}$,  they are related via the Bogoliubov transformation, which is 
\bq
\lb{4a5}
\nu_k=\alpha_k \mu_k+\beta_k \mu^*_k,
\eq
with $|\alpha_k|^2-|\beta_k|^2=1$ for any $k$. Since (3.5) is a linear equation and the  Bogoliubov coefficients are time-independent, the power spectra resulting from these two sets of initial states can be shown as 
\bq
\lb{4a6}
\mathcal P_{\nu_k}=\left(1+2|\beta_k|^2+2\mathrm{Re} \left(\alpha_k\beta^*_k\mu^2_k/|\mu_k|^2\right) \right)\mathcal P_{\mu_k}.
\eq
As is common in the literature, for a comparison with observations it is more convenient to provide the power spectrum of the comoving curvature perturbation ${\cal R}_k$, which is related to the Mukhanov-Sasaki variable by means of ${\cal R}_k=\nu_k/z$, with $z=a\dot \phi/H$. Its power spectrum then reads
\bq
\lb{4a7}
\mathcal P_{{\cal R}_k}=\frac{\mathcal P_{\nu_k}}{z^2}=\frac{k^3}{2\pi^2}\frac{|\nu_k|^2}{z^2}.
\eq
As usual, the power spectrum is evaluated at the end of inflation when all the relevant modes are well outside the Hubble horizon.
It should be noted that although the above formula can be used to generate the new power spectra from the already-existing ones, it is only applicable to the regimes where  $W^{(2)}_k$ or $W^{(4)}_k$ remains a real number at the initial time, which equivalently requires $k^2+s\ge 0$ for $W^{(2)}_k$ and $f(s,k)\ge 0$  for $W^{(4)}_k$. As the effective mass $s$ is generally a function of time, the validity regime of (\ref{4a6})  changes when the initial states are imposed at different initial times.

\subsection{Comparison of the power spectra among loop cosmological models in the hybrid approach }

In this subsection, we  compare the scalar power spectra in the three loop cosmological models from the effective dynamics of  the hybrid approach.  The difference among the three models  mainly originates from  the different quantizations of the  gravitational sector  of the classical Hamiltonian constraint in the spatially flat universe. As a result, although the Mukhanov-Sasaki  equations in these models take the same form given in (\ref{2a10}), the explicit form of the time-dependent mass $s$ and the evolution of the background quantities, such as the scale factor and the Hubble rate, satisfy their respective Hamilton's equations in each model. In order to obtain the primordial scalar power spectrum in each model from numerical simulations, one needs to first fix the background dynamics. As discussed in the last subsection, the initial conditions for the background dynamics are chosen at the bounce.  The parameter space is one dimensional which is determined by the value of the scalar field and the sign of its velocity. In order to facilitate comparison of the three models,  the initial conditions for the background are chosen so that the number of the inflationary e-foldings are the same which is fixed to be $66.8$ in all three models. Moreover, the initial values of the inflaton field are chosen at the left wing of the Starobinsky potential with a positive velocity. Under these conditions, the initial values of the inflaton field in LQC, mLQC-I and mLQC-II are  given respectively  by 
\bq
\lb{4b1}
\phi_\mathrm{LQC}=-1.44,\quad \quad \phi_{{\scriptscriptstyle{\mathrm{I}}}}=-1.32, \quad \quad \phi_{{\scriptscriptstyle{\mathrm{II}}}}=-1.55.
\eq
Under these initial conditions, we find the number of the pre-inflationary e-foldings which is counted from the bounce to the onset of inflation, in each model, turns out to be, respectively
\bq
\lb{4b2}
N^\mathrm{LQC}_\mathrm{pre}=4.86, \quad \quad  N^{{\scriptscriptstyle{\mathrm{I}}}}_\mathrm{pre}=4.62,\quad \quad  N^{{\scriptscriptstyle{\mathrm{II}}}}_\mathrm{pre}=5.10. 
\eq
In addition,  when the pivot mode crosses the horizon during the slow-roll phase, its co-moving wavenumber in three models are found to be 
\bq
\lb{comoving}
k^\mathrm{LQC}_*=5.15, \quad \quad  k^{{\scriptscriptstyle{\mathrm{I}}}}_*=4.05,\quad \quad  k^{{\scriptscriptstyle{\mathrm{II}}}}_*=6.56. 
\eq
Therefore, the observable window which is about $k/k_*\in(0.1,1000)$ in the three models is slightly shifted when they have the same inflationary e-foldings. Of course, one can fine tune the initial conditions so that  the  observable window is the same but inflationary e-foldings are different in the three models.

\begin{figure}
\includegraphics[width=8cm]{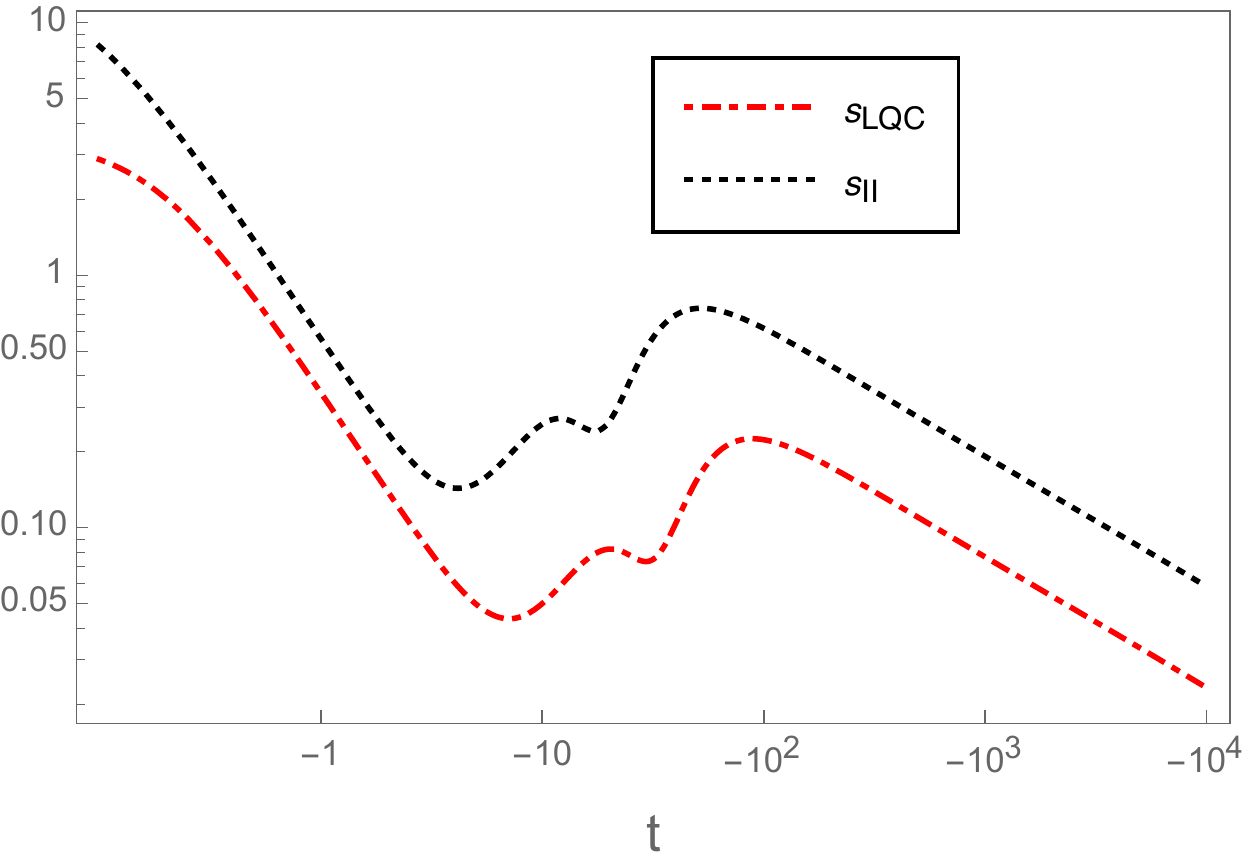}
\includegraphics[width=8cm]{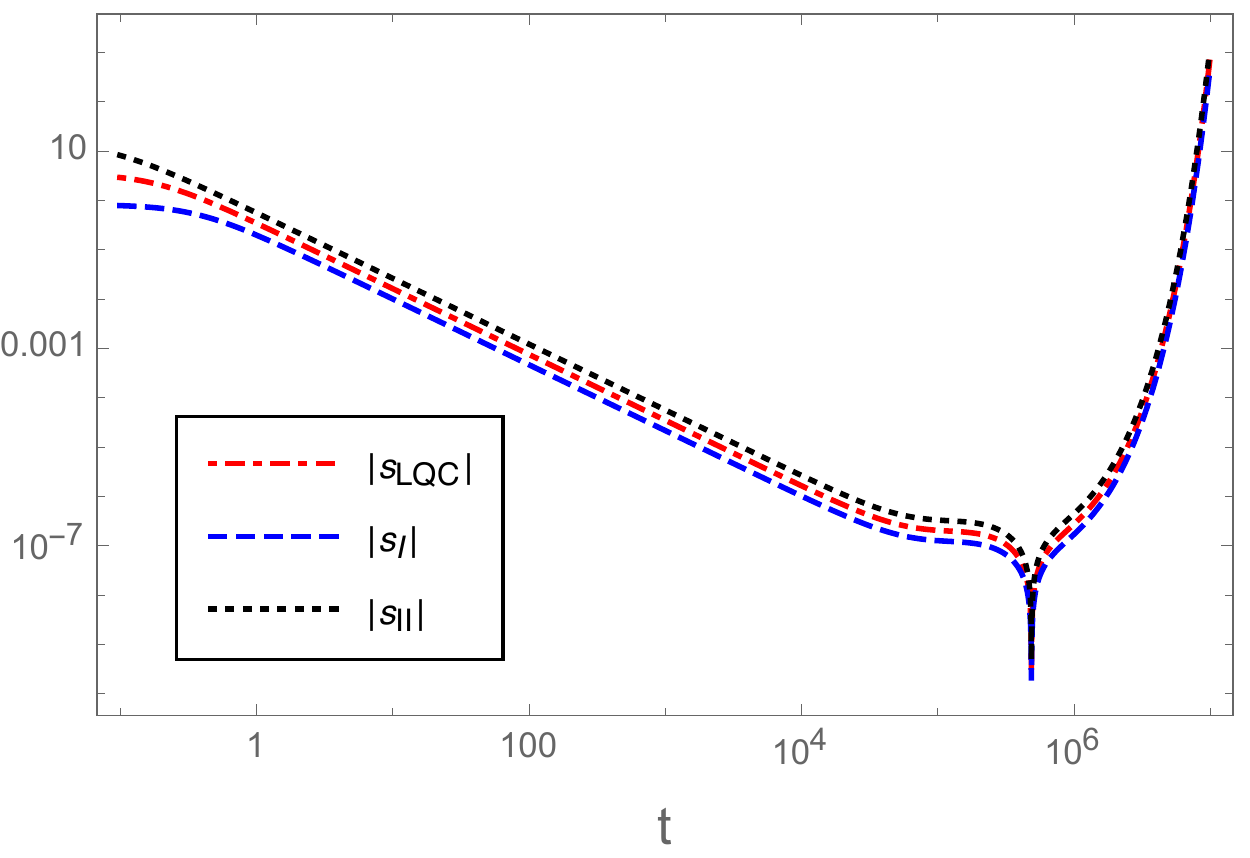}
\caption{The left panel compares the effective masses in LQC ($s_\mathrm{LQC}$) and mLQC-II ($s_{{\scriptscriptstyle{\mathrm{II}}}}$) from the hybrid approach in the contracting phase until the moment when the initial states are imposed. The right panel depicts the absolute value of the effective masses in LQC, mLQC-I ($s_{{\scriptscriptstyle{\mathrm{I}}}}$)  and mLQC-II until $t= 10^7 ~t_\mathrm{Pl}$. Right after the bounce, the effective masses take the positive values in all three models. During inflation, the effective masses change signs which produces the spikes in the right panel.} 
\label{f1}
\end{figure}

\begin{figure}
\includegraphics[width=8cm]{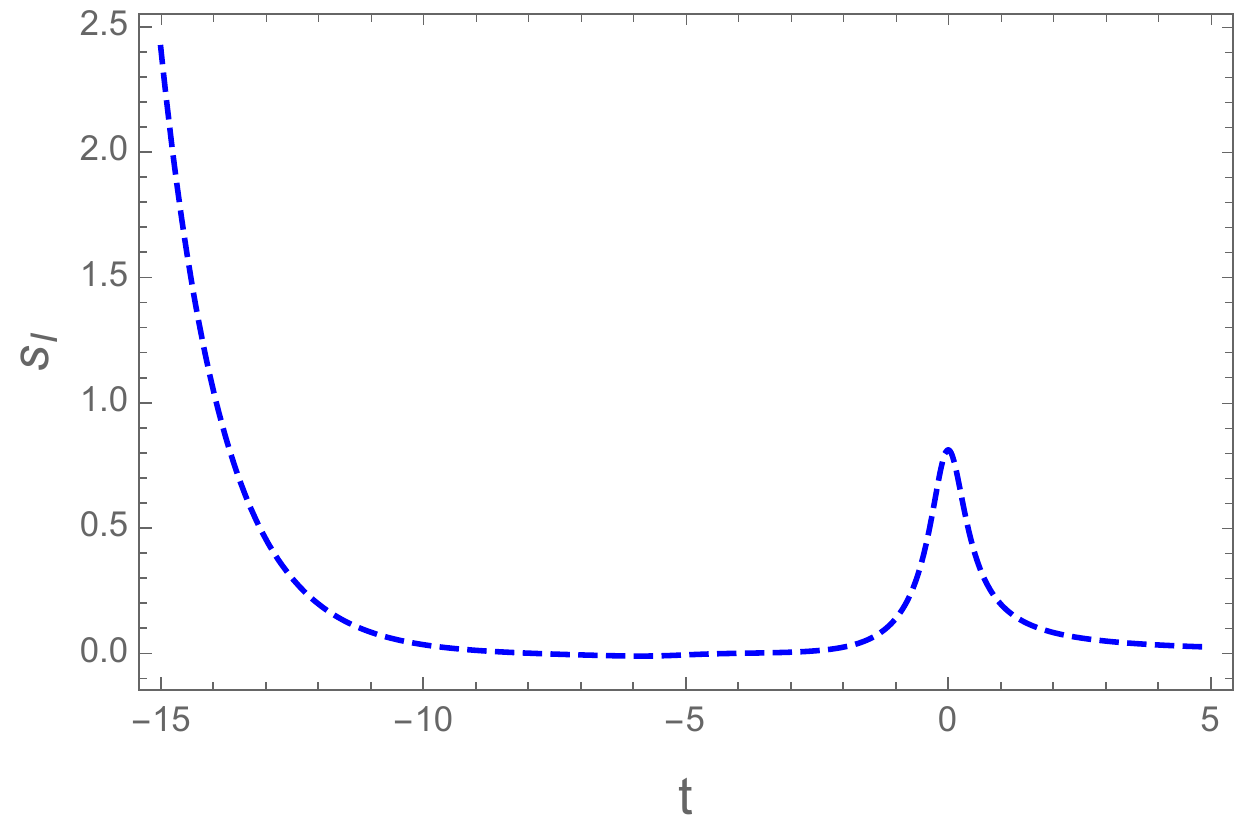}
\includegraphics[width=8cm]{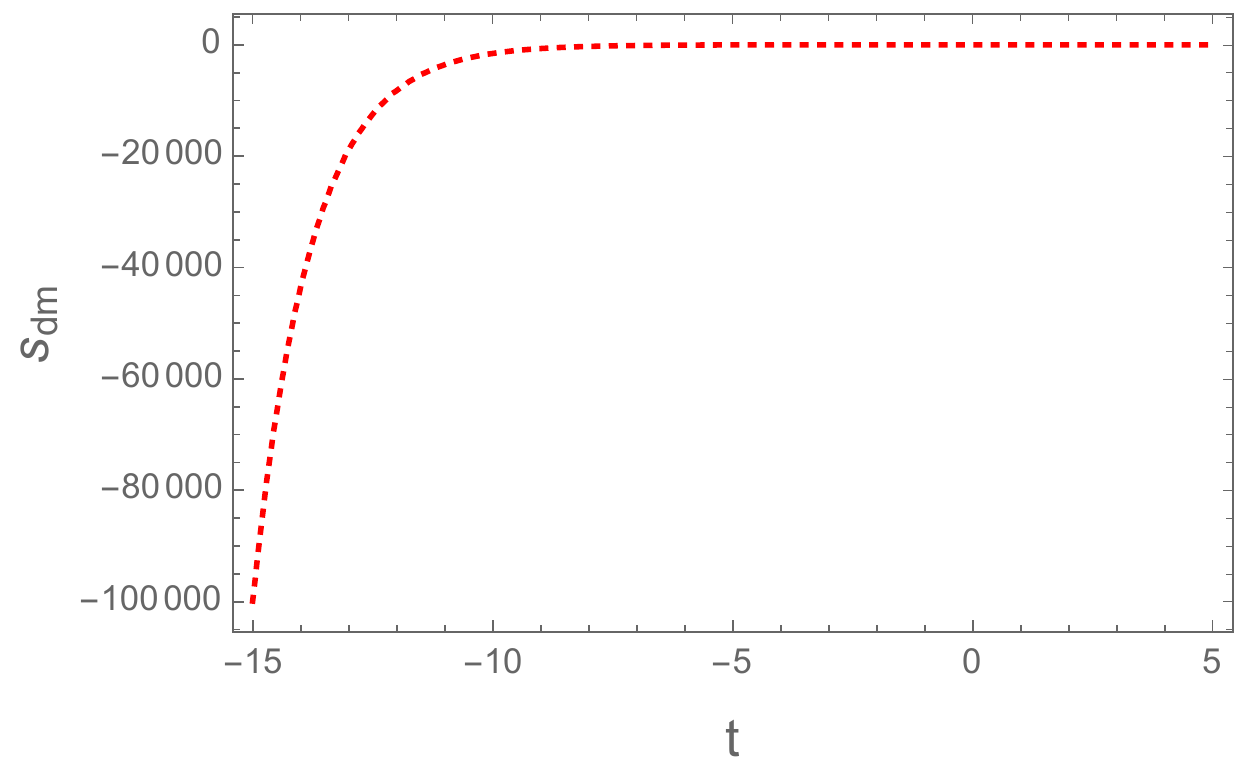}
\caption{In this figure, we compare the behavior of the effective masses in mLQC-I in the hybrid approach and the dressed metric approach. The left panel shows the effective mass $s_{{\scriptscriptstyle{\mathrm{I}}}}$ near the bounce in the hybrid approach while  the right panel depicts the effective mass $s_\mathrm{dm}$ in the dressed metric approach. Since in the contracting phase the universe quickly approaches the de Sitter space in the backward evolution, $s_\mathrm{dm}$  exponentially tends to negative infinity which is in contrast with the positive  $s_{{\scriptscriptstyle{\mathrm{I}}}}$.}
\label{f2}
\end{figure}

After fixing the background, one can then proceed to choose  the initial states for the scalar perturbations. These initial states are set in the contracting phase. For LQC and mLQC-II, we set the initial states at $t=-10^4~t_\mathrm{Pl}$ while for mLQC-I, the initial states are set at $t=-2$ where the spacetime is well approximated as being sourced by a positive cosmological constant. Different models  are mainly differentiated  by the effective masses  in the Mukhanov-Sasaki  equation, and we compare  these masses in the  three models in Fig. \ref{f1}. In the right panel of Fig. \ref{f1}, the absolute values of the  effective masses in LQC, mLQC-I/II are depicted in the expanding phase until $t=10^7~ t_\mathrm{Pl}$. Right after the bounce, the effective masses in all three models take positive values. During inflationary phase, the effective masses change their signs and thus produce the spikes in the figure. As can be seen from the figure, the behavior of the effective masses is qualitatively similar in these three models in the expanding phase while in the contracting phase, only LQC and mLQC-II have the qualitatively similar behavior of  effective masses. The behavior of the effective mass in mLQC-I is quite different from LQC and mLQC-II in the pre-bounce phase,  and is shown separately in the left panel of Fig. \ref{f2} where it is compared with the effective mass in the dressed metric approach in the right panel.  

Note that in the dressed metric approach, the  Mukhanov-Sasaki  equation in the quasi de Sitter contracting phase of mLQC-I takes the form 
\bq
\nu^{\prime \prime}_k+\left(k^2-\frac{2}{\eta^2}\right)\nu_k=0,
\eq
so the corresponding effective mass is given by 
\bq
\lb{4b3}
s_\mathrm{dm}= -\frac{2}{\eta^2},
\eq
where the prime denotes the derivatives with respect to the conformal time and the contributions from the inflationary potential is ignored as it is much smaller when compared with the contributions from the Planck-scale curvature near the bounce \cite{lsw2020}. One can  immediately find the difference between  the effective masses in the two different approaches. In the dressed metric approach, the effective mass takes the negative values and increases exponentially in magnitude  during the backward evolution in the contracting phase. Then,  the following initial state of the linear perturbations is chosen
\bq
\lb{bd}
\nu_k=\frac{e^{-ik\eta}}{\sqrt{2k}}\left(1-\frac{i}{k\eta}\right).
\eq
It should be noted that the modes in the infrared and intermediate regimes are outside the Hubble horizon initially, which indicates  $k\eta\ll1$ at the time when the initial states are imposed. As a result, the second term in the parenthesis of (\ref{bd}) can not be ignored for those modes. Only the modes in the ultraviolet regime are initially inside the Hubble horizon and hence their initial states coincide with  the zeroth order adiabatic states. On the other hand, in the hybrid approach, the property of the effective mass turns out to be quite different. The effective mass now takes positive values and  increases in a non-exponential way. Consequently, in the hybrid approach,  all the relevant modes are inside the Hubble horizon at the initial time. For this reason, one would expect the difference between the power spectra from two approaches will mainly occur in the infrared and intermediate regimes. However, we  will use the second order adiabatic states for the numerical simulations of the power spectrum in the hybrid approach. The use of other initial states, like the zeroth or fourth order adiabatic state, will not qualitatively change our results.  
\begin{figure}
\includegraphics[width=8cm]{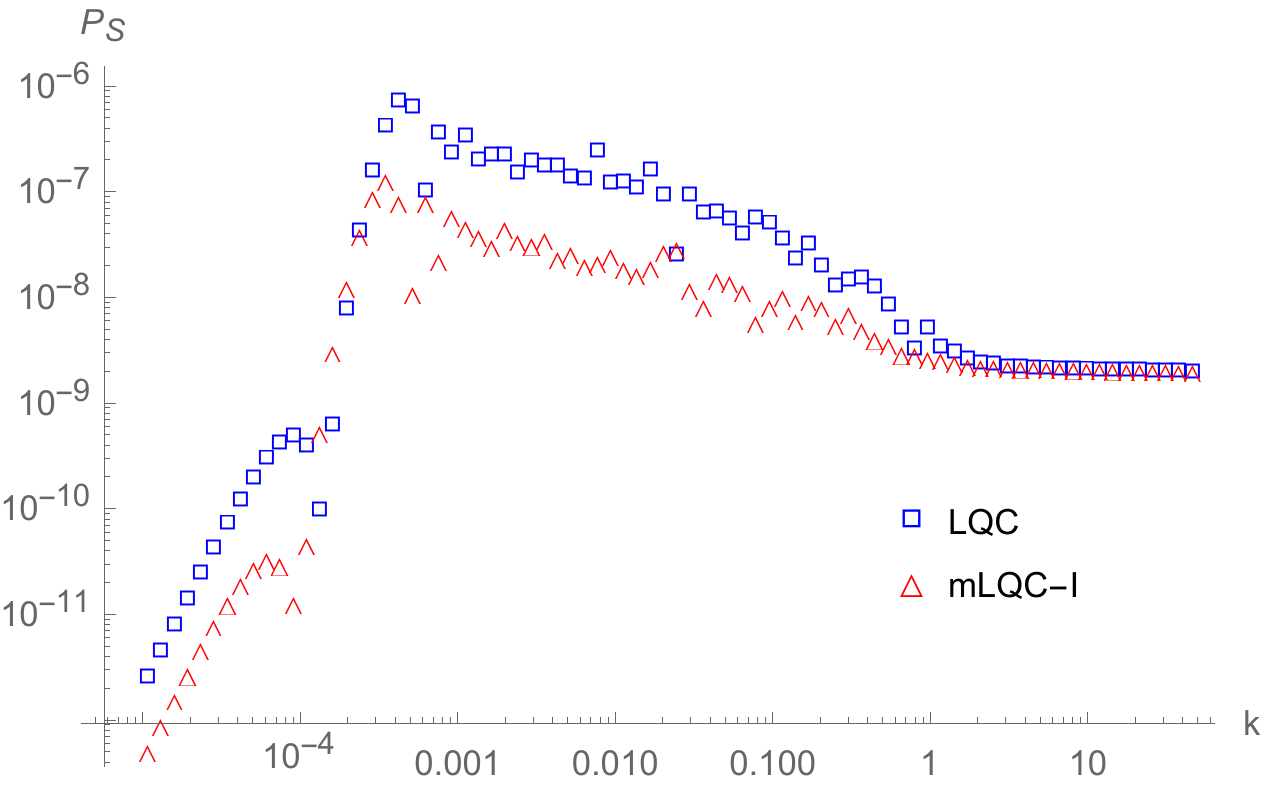}
\includegraphics[width=8cm]{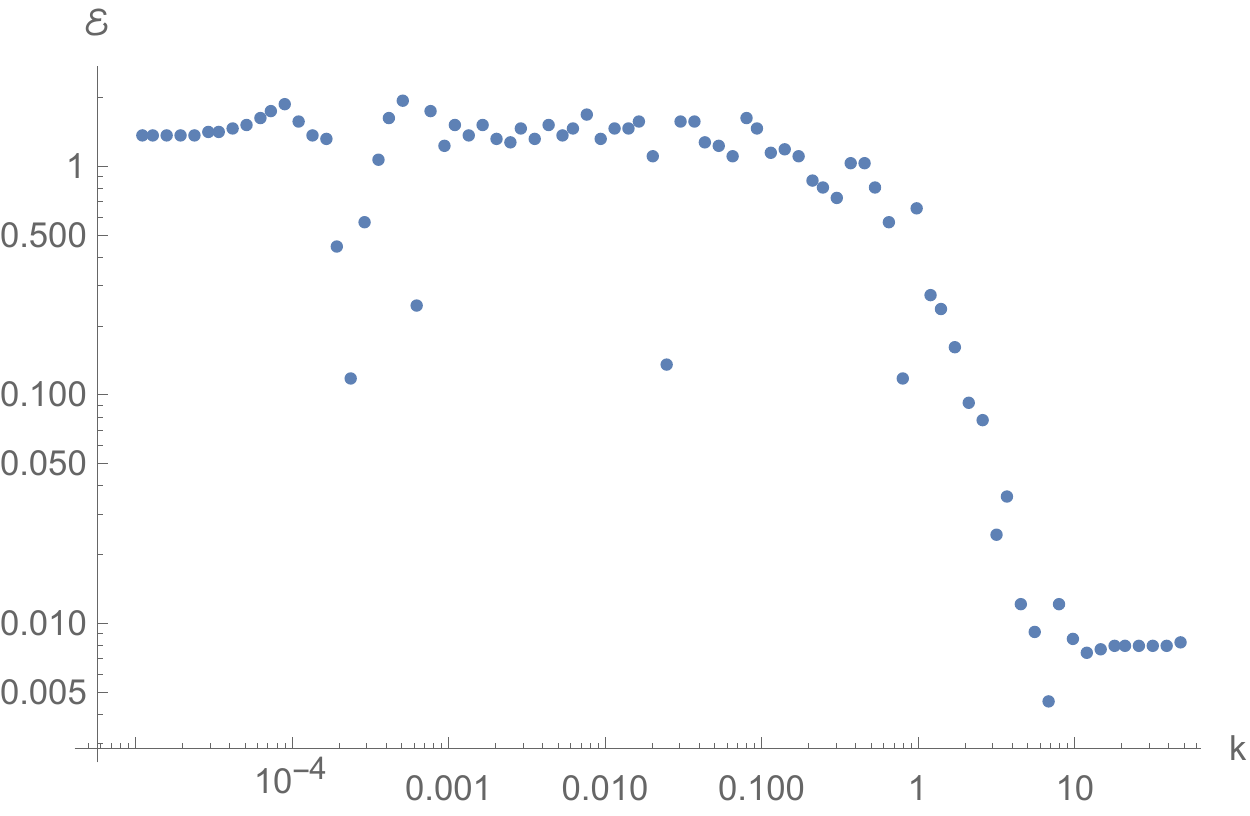}
\caption{ In this figure, we compare the scalar power spectra  for the modes $k\in(10^{-5},50)$ in LQC (blue square) and mLQC-I (red triangle) from the hybrid approach when the initial states are chosen to be the second order adiabatic states and imposed in the contracting phase.  The right panel shows the relative difference of the two power spectra defined in (\ref{4b4}). }
\label{f3}
\end{figure}

\begin{figure}
\includegraphics[width=8cm]{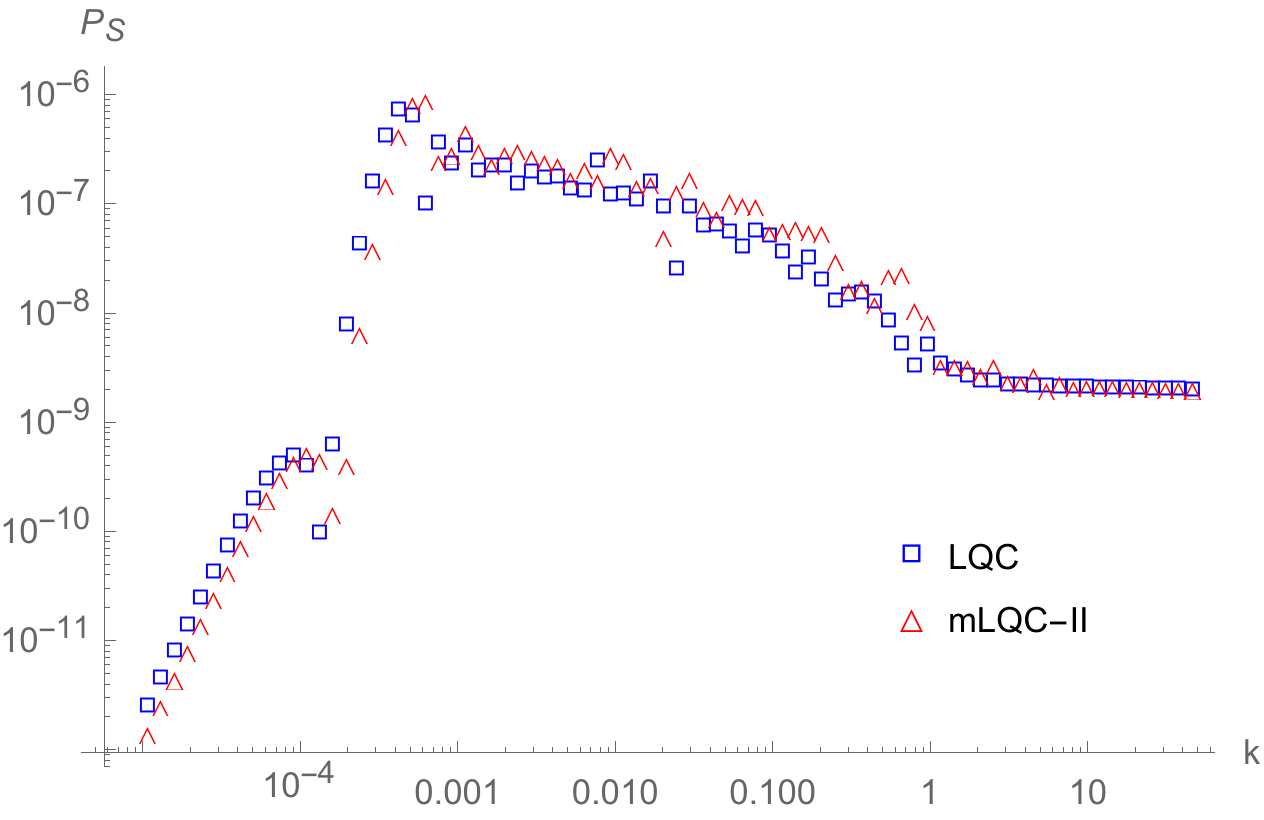}
\includegraphics[width=8cm]{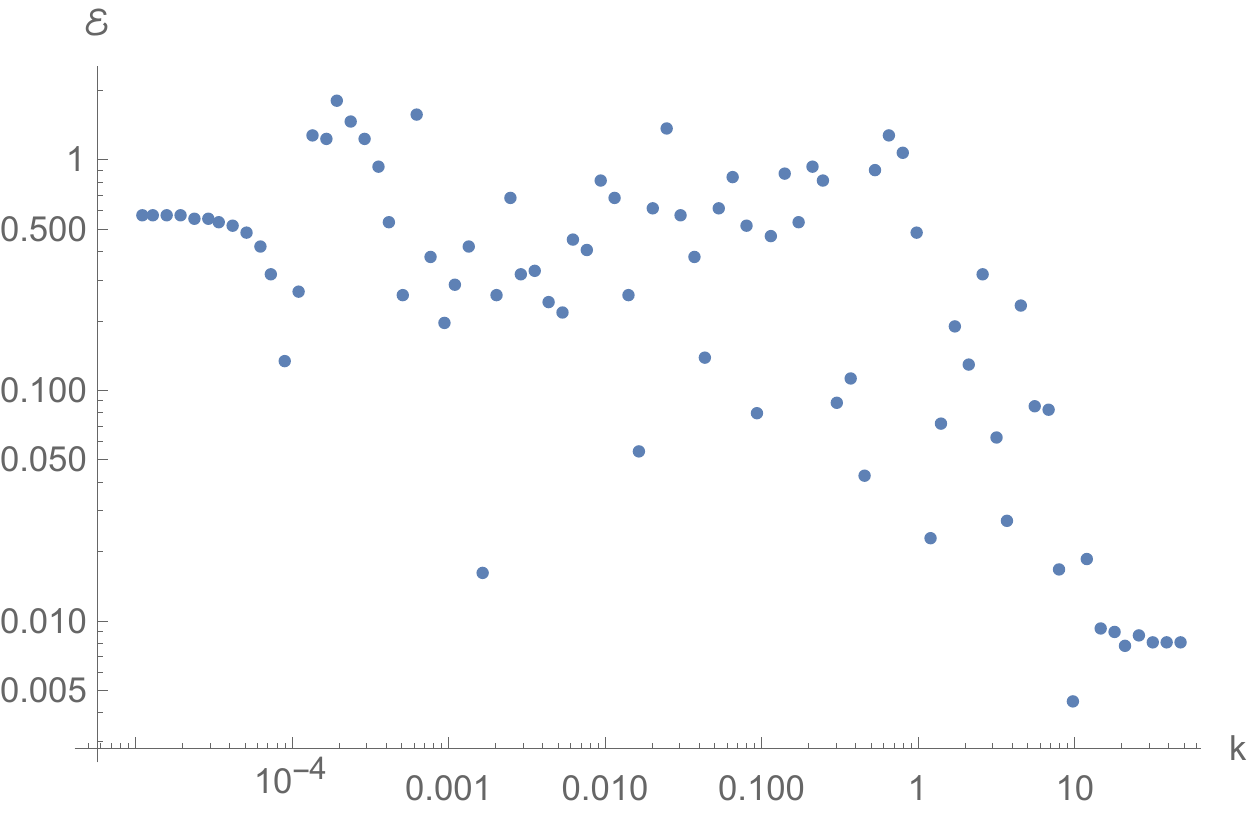}
\caption{The scalar power spectra from the hybrid approach are depicted for LQC (blue square) and mLQC-II (red triangle) with the initial states imposed at $t=-10^4$ in both models. These initial states are the second order adiabatic states. The right panel shows the relative difference between the two models.}
\label{f4}
\end{figure}

\begin{figure}
\includegraphics[width=8cm]{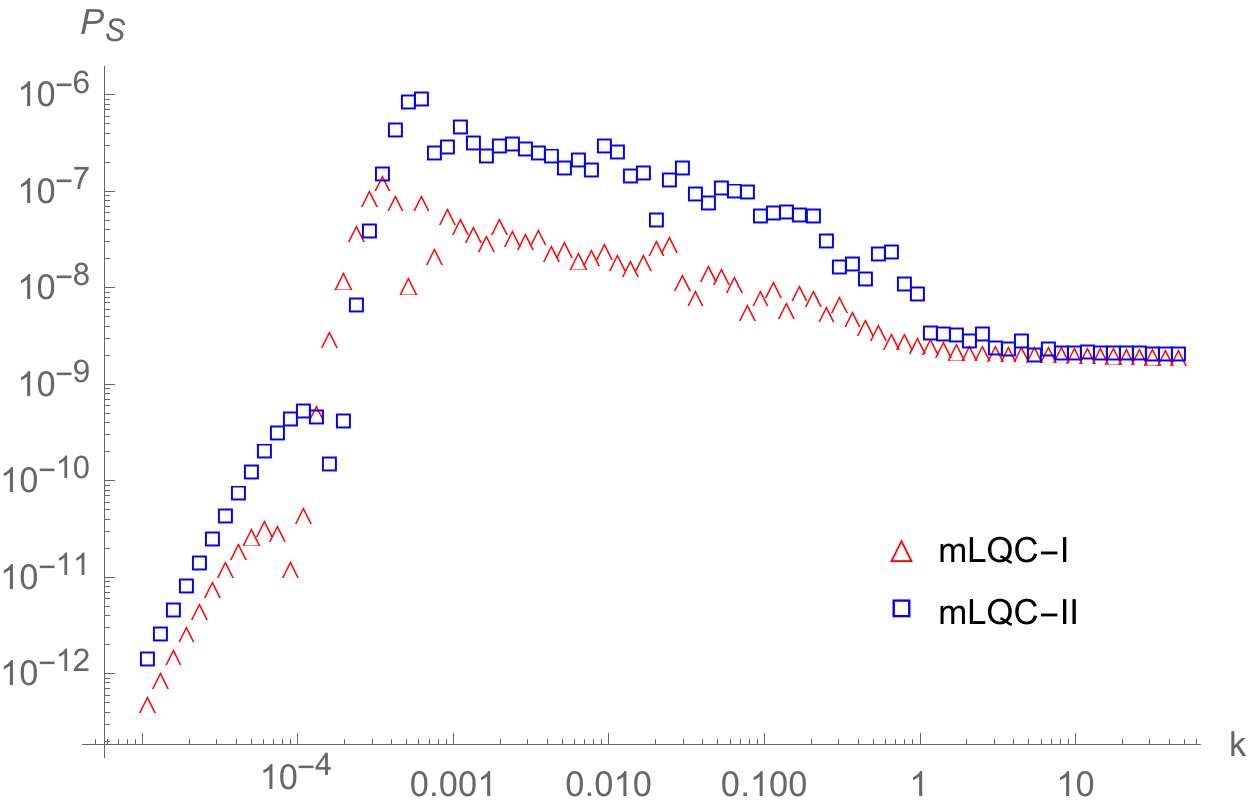}
\includegraphics[width=8cm]{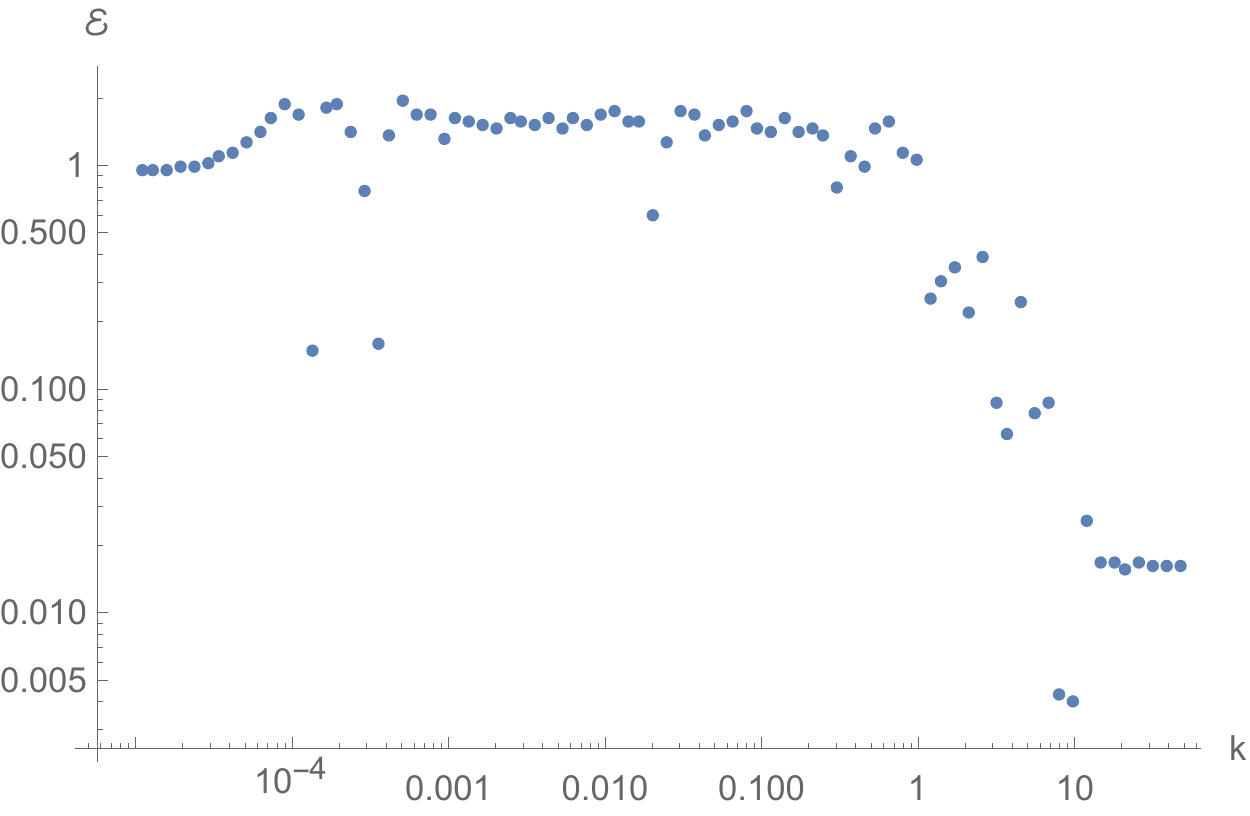}
\caption{We compare the scalar power spectra (left panel) in mLQC-I (red triangle) and mLQC-II  (blue square) from the hybrid  approach and show the relative difference between the power spectra (right). The initial states are imposed at $t=-10^4$ in mLQC-II and  $t=-2$ in mLQC-I.  }
\label{f5}
\end{figure}

Our final results on the power spectra  are presented and compared in Figs. \ref{f3}-\ref{f5}. 
In Fig. \ref{f3}, the scalar power spectra in LQC and mLQC-I are compared in the  range of the co-moving wavenumber $k\in (10^{-5}, 50)$.  The power spectrum can still be divided into three distinctive regimes: the suppressed infrared regime for $k\approx (10^{-5}, 10^{-4})$, the amplified oscillatory regime for $k\approx(10^{-4},1)$ and the scale invariant regime for $k\approx (1, 50)$. Although the power spectra in LQC and mLQC-I have  the similar qualitative behavior throughout the considered range of the wavenumber, their quantitative difference can be seen from the right panel of Fig. \ref{f3} in which the relative difference $\mathcal E$ is shown. For any two quantities $\mathcal Q_1$ and $\mathcal Q_2$, the relative difference $\mathcal E$ is defined by 
\bq
\lb{4b4}
\mathcal E= 2\frac{|\mathcal Q_1-\mathcal Q_2|}{|\mathcal Q_1+\mathcal Q_2|}. 
\eq
In the infrared and  oscillatory regimes, the relative difference can reach as  large as $100\%$ while the difference reduces  to less than $1\%$ in the scale invariant regime. This is primarily because LQC and mLQC-I  have the same classical limit in the expanding phase,  and   as shown in Fig. \ref{f1}, the effective masses in both approaches also tend to the same  value in the  inflationary phase. It is also remarkable to note that in the infrared and  oscillatory regimes, the power spectrum in mLQC-I is suppressed as compared with its counterpart in LQC. This is a very unique feature manifest only  in the hybrid approach. In the dressed metric approach, the power spectrum in mLQC-I is largely amplified in the  infrared regime where its magnitude is as large as of the Planck scale \cite{IA19,lsw2020}. The main reason that causes this seemingly contradictory behavior of the power spectrum in mLQC-I in both approaches lies in the distinctive behavior of the effective masses in the two approaches as depicted in Fig. \ref{f2} and the corresponding choices of the initial states in the contracting phase. In Fig. \ref{f4}, the power spectra in LQC and mQLC-II are compared. As expected from the similarity of the effective masses in these two models, the relative difference between the power spectra of these two models are smaller than the relative difference between LQC and mLQC-I. In the  infrared  regime, the relative difference is around $50\%$. The relative difference  in the oscillatory regime also oscillates as in this regime the oscillations of the power spectrum in LQC and mLQC-II are in general  out of phase. The comparison between the power spectra from mLQC-I and mLQC-II are presented in Fig.  \ref{f5} where we find that  a large relative difference (more than $100\%$) is still present in the infrared and oscillatory regimes while the relative difference in the scale invariant regime is around $2\%$. 

In the above analysis, we have compared the scalar power spectrum from the three models when the initial conditions of the background dynamics are chosen at the bounce so that the numbers of the inflationary e-foldings turn out to be the same. This results in some difference in the co-moving  wavenumbers of the pivot mode at the horizon crossing in the three models as presented in (\ref{comoving}). Due to this, there is an overlap in the value of $k$ for the observable window. Hence, a comparison of the primordial power spectrum among the three different models can be reliably made. In principle, one can choose other initial conditions such that the wavenumbers of the pivot mode are exactly the same in the three models. This would imply the inflationary e-foldings would be different. Then the observable window corresponds to the same range of the co-moving wavenumbers in all three models.  For this set of initial conditions of the background and with the same initial states for the perturbations used  in Figs. \ref{f3}-\ref{f5}, we find very similar results as presented in those figures. 

Let us summarize the results from numerical simulations.  After fixing the initial conditions of the background dynamics and the initial states of the scalar perturbations for LQC, mLQC-I and mLQC-II in the hybrid approach we compared the effective masses and the resulting power spectra in these three models. We found a similar pattern of the power spectra from the three models which can be divided into three distinctive regimes. The maximum relative difference of the power spectra from different models are reached in the infrared and oscillatory regimes  while in the scale invariant regime, all three models predict a similar result which is consistent with the current CMB observations. It is to be emphasized that in the hybrid approach, the power spectrum in mLQC-I is suppressed in the infrared and oscillatory regimes which is in a striking contrast with the results from the dressed metric approach. This remarkable difference originates from the distinctive properties of the effective masses in these two approaches and reveals for the first time differences in predictions due to underlying construction in these two approaches.

\section{Conclusions}
\lb{summary}
\renewcommand{\theequation}{5.\arabic{equation}}\setcounter{equation}{0}

In this paper,  we discussed the effective dynamics of the hybrid approach in the modified loop cosmological models, namely mLQC-I and mLQC-II. For this purpose, we first briefly reviewed the effective dynamics of the hybrid approach in LQC, including the effective equations for the background dynamics and the gauge invariant perturbations. 
An important step for deriving the Mukhanov-Sasaki equation  in LQC is the specification of the operator $\hat \Lambda$ which is well-defined in the subspaces $\mathcal H^\pm_\epsilon$ selected by the homogeneous scalar constraint. Following the same strategy,  we  specified the operator analogs to $\hat \Lambda$ and their effective counterparts in mLQC-I/II. It turns out that  the Mukhanov-Sasaki equation takes the same form in these two models as in LQC, and the only difference lies in the effective masses which have distinct behavior in each model.

 In order to quantitatively study the difference in the power spectra of the three loop cosmological models, we then considered the Starobinsky inflation driven by a single scalar field and  found numerical solutions of the background and the perturbations under a representative set of initial conditions which makes the inflationary e-foldings equal in the three models. The initial states for the perturbations are chosen to be the second order adiabatic states and imposed in the contracting phase.  Under these conditions, we compared the effective masses and the scalar power spectra in LQC and mLQC-I/II. In the expanding phase, the effective masses are qualitatively similar in the three models, they are initially positive valued and deceasing in the pre-inflationary stage. Later, the effective masses change sign during inflation and their magnitudes keep increasing until the end of the inflation. Since the square of  the comoving Hubble horizon is  given by the negative of the inverse of the effective mass, the behavior of the effective masses in the three models is consistent with the deceasing comoving Hubble horizon during inflation. Based on the numerical solutions of the background dynamics, we find in the contracting phase the effective masses in LQC and mLQC-II have similar properties, both of them tend to decrease in the backward evolution from the bounce while the effective mass in mLQC-I has qualitative different behavior. Initially, the effective mass in mLQC-I is decreasing in the backward evolution from the bounce. When the background spacetime becomes the de Sitter space, the effective mass tends to climb up. We find that in the hybrid approach, the change rate of the effective mass is much slower than that in the dressed metric approach, and most importantly the effective masses in these two approaches also have opposite signs.  As a result, in the dressed metric approach, only the ultraviolet modes are inside the Hubble horizon when the initial states are imposed while in the hybrid approach, all the relevant modes are well inside the horizon. 
 This is a key difference between the two approaches. 
 
 The resulting power spectra in LQC and mLQC-I/II also assume the similar patterns  with three distinctive regimes: the infrared regime, the oscillatory regime and the ultraviolet regime. The magnitudes of the power spectra in the three models are comparable in all three regimes. Quantitatively,  more diversities are present in the infrared and oscillatory regimes than in the ultraviolet regime. The relative difference of the power spectra can be as large as $100\%$ between LQC/mLQC-II and mLQC-I and  $50\%$ between LQC and mLQC-II in the  former regimes while in the ultraviolet regime, all three models predict the  scale invariant power spectra which are consistent with the observations within the numerical errors.  Furthermore, the magnitude of the power spectrum in mLQC-I is suppressed in the infrared and oscillatory regimes as compared with the power spectra in LQC and mLQC-II. This behavior is very distinct if compared with the results from the dressed metric approach in \cite{lsw2020} where a Planck scale magnitude of the power spectrum in the same regimes is found in mLQC-I.  The difference between the two different approaches for mLQC-I originates from the distinctive properties of the effective masses in the two approaches. It is remarkable since this is for the first time in loop cosmology the dressed and hybrid approaches yield significantly different predictions in the power spectrum. 
 
We would like to emphasize that our results are robust with respect to the choices of the initial conditions  and the initial adiabatic states. Although the initial volume is set to unity, we have to note that only the holonomy corrections are considered in the effective dynamics. As the equations of the motion are invariant under the rescaling of the volume, it is convenient to set the initial volume to unity at the bounce. Any rescaling of the initial volume is equivalent to rescale the comoving wavenumbers and thus translate the power spectrum  as  a whole (as it would happen in standard GR). The different choices of the initial values of the scalar field can change the e-folding from the bounce to the horizon exit of the pivot mode and thus move the observable windows in the power spectrum. The different choices of the adiabatic states can be related via the Bogoliubov transformation, and the resulting averaged power spectra will differ by a constant determined by the initial states. As a result, the relative difference of the power spectra from different models will not change by specifying a different initial state instead of the second order adiabatic states. However, the absolute magnitude of the power spectra in the infrared regime does depend on the initial states as shown in \cite{bo2016}.  Since our main purpose is to study the difference between LQC and mLQC-I/II, we find it sufficient to show for the considered set of initial conditions. Moreover, our result that power spectrum in mLQC-I is significantly different in the dressed and hybrid approaches is also independent of the choices discussed above since it is tied to the effective masses which turn out to be significantly different in both of the approaches. 

Finally, we  conclude with following remarks.  Although from our numerical analysis we found that the power spectra from both, the hybrid and the dressed metric approaches for mLQC-I are only different in the infrared and oscillatory regimes and consistent with the CMB observations in the ultraviolet regime  at the level of the linear perturbations,  it is essential to consider the non-Gaussianity in mLQC-I to fully compare the differences between two approaches in the observable regime as the magnitude of the power spectrum  from the dressed metric approach are of the Planck scale in the infrared and oscillatory regimes. Therefore,  the perturbations with the Planck-scale magnitude in the long wavelength modes are quite likely to affect the magnitude of the power spectrum of the  short wavelength modes through the interactions between these modes. Unlike the dressed metric approach,  the small magnitude of the power spectrum in the hybrid approach throughout the whole spectrum justifies its application to mLQC-I at the level of the linear perturbations. It also implies that  at the linear order, the hybrid approach is well suited to the different quantizations of the classical Hamiltonian constraint in a spatially flat FLRW universe in LQC.

\section*{Acknowledgements}
We thank Guillermo Mena Marug\'an for comments on this manuscript. 
B.F.L. and P.S. are supported by NSF grant PHY-1454832. J.O. acknowledges the Operative Program FEDER 2014-2020 and Consejer\'ia de Econom\'ia y Conocimiento de la Junta de Andaluc\'ia.
 A.W. is supported in part by the National Natural Science Foundation of China (NNSFC) 
with the Grants Nos. 11675145 and 11975203.


\begin{thebibliography}{99}

   \bibitem{ag1981}  A. H. Guth, {\em Inflationary universe: A possible solution to the horizon and flatness problems}, Phys. Rev. D{\bf 23}, 347 (1981).
   
      
      \bibitem{lss}    D. H. Lyth and A. R. Liddle,  {\em The Primordial Density Perturbation: Cosmology, Inflation and the Origin of Structure} (Cambridge University Press, London, 2009).
      
       \bibitem{BV94} A. Borde and A. Vilenkin, {\em Eternal inflation and the initial singularity}, Phys. Rev. Lett. {\bf 72}, 3305 (1994); A. Borde, A. H. Guth,  A. Vilenkin, {\em Inflationary spacetimes are incomplete in past directions}, Phys. Rev. Lett. {\bf 90}, 151301 (2003).
       

   \bibitem{review2} A.~Ashtekar and P.~Singh, {\em Loop quantum cosmology: a status report},  Class. Quant. Grav.  {\bf 28}, 213001 (2011).
   
    \bibitem{aps1} A.~Ashtekar, T.~Pawlowski and P.~Singh, {\em Quantum nature of the big bang},  Phys. Rev. Lett.  {\bf 96}, 141301 (2006).
    
      \bibitem{aps2}   A.~Ashtekar, T.~Pawlowski and P.~Singh, {\em Quantum nature of the big bang: an analytical and numerical investigation. I.,} Phys.\ Rev.\ D{\bf 73}, 124038 (2006).
  
 \bibitem{aps3} A. Ashtekar, T. Pawlowski and P. Singh, {\em Quantum nature of the big bang: improved dynamics}, Phys. Rev. D{\bf 74}, 084003 (2006).
 
   \bibitem{acs2010}   A. Ashtekar, A. Corichi. and P. Singh,  {\em Robustness of key features of loop quantum cosmology}, Phys. Rev. D{\bf 77}, 024046 (2010). 
   
    \bibitem{gls2020}  K. Giesel, B. F. Li  and P. Singh, {\em Towards a reduced phase space quantization in loop quantum cosmology with an inflationary potential}, {arXiv:2007.06597}.
    
         \bibitem{as2017}  I.~Agullo and P.~Singh, Loop Quantum Cosmology, in {\it{Loop Quantum Gravity: The First 30 Years}}, Eds: A. Ashtekar, J. Pullin, World Scientific (2017) arXiv:1612.01236.
 
 \bibitem{numlsu-1}   P.~Singh, {\em Glimpses of Space-Time Beyond the Singularities Using Supercomputers,} Comput.\ Sci.\ Eng.\  {\bf 20}, 26 (2018).

 
  \bibitem{numlsu-2} P.~Diener, B.~Gupt and P.~Singh,   {\em Numerical simulations of a loop quantum cosmos: robustness of the quantum bounce and the validity of effective dynamics}, Class.\ Quant.\ Grav.\  {\bf 31}, 105015 (2014).

  
 \bibitem{numlsu-3} P.~Diener, B.~Gupt, M.~Megevand and P.~Singh, {\em Numerical evolution of squeezed and non-Gaussian states in loop quantum cosmology}, Class.\ Quant.\ Grav.\  {\bf 31}, 165006 (2014).

  
 \bibitem{numlsu-4} P.~Diener, A.~Joe, M.~Megevand and P.~Singh,   {\em Numerical simulations of loop quantum Bianchi-I spacetimes}, Class.\ Quant.\ Grav.\  {\bf 34},  094004 (2017).
 

  
   
  
   
 
 
  
  
      

     
  
 
  \bibitem{YDM09}  J. Yang, Y. Ding and Y. Ma, {\em Alternative quantization of the Hamiltonian in loop quantum cosmology II: Including the Lorentz term}, Phys. Lett.  B{\bf 682} (2009) 1. 

\bibitem{DL17} A. Dapor and K. Liegener, {\em Cosmological effective Hamiltonian from full loop quantum gravity dynamics}, Phys. Lett. B{\bf 785} (2018) 506.


\bibitem{mehdi} Mehdi Assanioussi, Andrea Dapor, Klaus Liegener, and Tomasz Pawłowski, {\em Emergent de Sitter epoch of the quantum cosmos from loop quantum cosmology}, Phys. Rev. Lett. {\bf 121} (2018) 081303. 


\bibitem{lsw2018} B. F. Li, P.~Singh, A. Wang,  {\em Towards cosmological dynamics from loop quantum gravity}, Phys. Rev. D{\bf 97}, 084029 (2018).

\bibitem{ss6} 
  S.~Saini and P.~Singh,
  {\em Von Neumann stability of modified loop quantum cosmologies,}
  Class.\ Quant.\ Grav.\  {\bf 36}, 105010 (2019).


\bibitem{ss5} 
  S.~Saini and P.~Singh, {\em Generic absence of strong singularities and geodesic completeness in modified loop quantum cosmologies,}
  Class.\ Quant.\ Grav.\  {\bf 36}, 105014 (2019).
  

    \bibitem{lsw2018b} B. F. Li, P.~Singh, A. Wang, {\em Qualitative dynamics and inflationary attractors in loop cosmology}, Phys. Rev. D{\bf 98}, 066016 (2018).

    \bibitem{lsw2019} B. F. Li, P.~Singh, A. Wang, {\em Genericness of pre-inflationary dynamics and probability of the desired slow-roll inflation in modified loop quantum cosmologies}, Phys. Rev. D{\bf 100}, 063513 (2019).
    
    \bibitem{djs} N.~Dadhich, A.~Joe and P.~Singh, {\em 
 Emergence of the product of constant curvature spaces in loop quantum cosmology,}
  Class.\ Quant.\ Grav.\  {\bf 32},  185006 (2015).
  
  \bibitem{pert-old}  S.~Tsujikawa, P.~Singh and R.~Maartens, {\em 
  Loop quantum gravity effects on inflation and the CMB,}
  Class.\ Quant.\ Grav.\  {\bf 21}, 5767 (2004); 
  G.~M.~Hossain,
  {\em Primordial density perturbation in effective loop quantum cosmology,}
  Class.\ Quant.\ Grav.\  {\bf 22}, 2511 (2005); X. Zhang and Y. Ling, {\em Inflationary universe in loop quantum cosmology}, JCAP 08, 012 (2007); M.~Bojowald, H.~Hernandez, M.~Kagan, P.~Singh and A.~Skirzewski, {\em Formation and evolution of structure in loop cosmology,}
  Phys.\ Rev.\ Lett.\  {\bf 98}, 031301 (2007); J.~Magueijo and P.~Singh, {\em 
  Thermal fluctuations in loop cosmology,}
  Phys.\ Rev.\ D{\bf 76}, 023510 (2007);  M.~Shimano and T.~Harada, {\em Observational constraints on a power spectrum from super-inflation in loop quantum cosmology},
  Phys.\ Rev.\ D{\bf 80}, 063538 (2009); J.~Grain and A.~Barrau,
  {\em Cosmological footprints of loop quantum gravity,}
  Phys.\ Rev.\ Lett.\  {\bf 102}, 081301 (2009); J.~Mielczarek,
 {\em Possible observational effects of loop quantum cosmology,}
  Phys.\ Rev.\ D{\bf 81}, 063503 (2010).
    
     \bibitem{bhks2008} M. Bojowald, G. M. Hossain, M. Kagan, S. Shankaranarayanan, {\em Anomaly freedom in perturbative loop quantum gravity},  Phys. Rev. D{\bf 78}, 063547 (2008).   
           
 \bibitem{cbgv2012} T. Cailleteau, A. Barrau, J. Grain, F. Vidotto, {\em Consistency of holonomy-corrected scalar, vector and tensor perturbations in loop quantum cosmology},  Phys. Rev. D{\bf 86}, 087301 (2012).  
                
  \bibitem{cmbg2012} T. Cailleteau, J. Mielczarek, A. Barrau, J. Grain, {\em Anomaly-free scalar perturbations with holonomy corrections in loop quantum cosmology}, Class.\ Quant.\ Grav.\  {\bf 29}, 095010 (2012).

\bibitem{wilson2017}E.~Wilson-Ewing,  {\em Testing loop quantum cosmology}, Comptes Rendus Physique {\bf 18}, 207 (2017). 

       
   \bibitem{aan2012}   I. Agullo, A. Ashtekar and W. Nelson, {\em Quantum gravity extension of the inflationary scenario}, Phys. Rev. Lett. {\bf 109}, 251301 (2012).
   
\bibitem{aan2013}   I. Agullo, A. Ashtekar and W. Nelson, {\em Extension of the quantum theory of cosmological perturbations to the Planck era}, Phys. Rev. D{\bf 87}, 043507 (2013).

\bibitem{aan2013b}  I. Agullo, A. Ashtekar and W. Nelson, {\em The pre-inflationary dynamics of loop quantum cosmology: confronting quantum gravity with observations}, Class. Quant. Grav. {\bf30}, 085014 (2013).


\bibitem{mm2012} M. Fern\'andez-M\'endez, G. A. Mena Marug\'an and J. Olmedo,  {\em Hybrid quantization of an inflationary universe}, Phys. Rev. D{\bf 86}, 024003 (2012).
  
\bibitem{mm2013} M. Fern\'andez-M\'endez, G. A. Mena Marug\'an and J. Olmedo,  {\em Hybrid quantization of an inflationary model: The flat case}, Phys. Rev. D{\bf 88}, 044013 (2013).

\bibitem{gmmo2014} L. Castell\'o Gomar, M. Fern\'andez-M\'endez, G. A. Mena Marugan and J. Olmedo, {\em Cosmological perturbations in hybrid loop quantum cosmology: Mukhanov-Sasaki variables}, Phys. Rev. D{\bf 90}, 064015 (2014).
  
\bibitem{gbm2015} L. Castell\'o Gomar, M. Mart\'in-Benito, G. A. Mena Marug\'an, {\em Gauge-invariant perturbations in hybrid quantum cosmology}, JCAP 1506 (2015) 045.

\bibitem{mo2016}   F. Ben\'itez Mart\'inez and J.~Olmedo, {\em Primordial tensor modes of the early Universe,} Phys.\ Rev.\ D{\bf 93},  124008 (2016).  

\bibitem{b1-lett}  I. Agullo, J. Olmedo and V. Sreenath, {\em Predictions for the cosmic microwave background from an anisotropic quantum bounce}, Phys. Rev. Lett. {\bf 124}, 251301 (2020).

\bibitem{b1-long}  I. Agullo, J. Olmedo and V. Sreenath, {\em Observational consequences of Bianchi I spacetimes in loop quantum cosmology}, arxiv: 2006.01883 (2020).

  
    \bibitem{d1} A.~Ashtekar and B.~Gupt, {\em Quantum gravity in the sky: Interplay between fundamental theory and observations,}
  Class.\ Quant.\ Grav.\  {\bf 34}, 014002 (2017).
  

\bibitem{d2} I.~Agullo, A.~Ashtekar and B.~Gupt, {\em Phenomenology with fluctuating quantum geometries in loop quantum cosmology,}
  Class.\ Quant.\ Grav.\  {\bf 34},  074003 (2017).


\bibitem{bo2016}  D.~Mart\'in de Blas and J.~Olmedo, {\em Primordial power spectra for scalar perturbations in loop quantum cosmology,}
  JCAP {\bf 1606}, 029 (2016).
  

  \bibitem{gbmo2016} 
  L. Castell\'o Gomar, D.~Mart\'in de Blas, G. A. Mena Marug\'an  and J.~Olmedo, {\em Hybrid loop quantum cosmology and predictions for the cosmic microwave background,}
  Phys.\ Rev.\ D{\bf 96},  103528 (2017).  

  \bibitem{tao2017}   T. Zhu, A. Wang, G. Cleaver, K. Kirsten, and Q. Sheng, {\em Pre-inflationary universe in loop quantum cosmology}, Phys. Rev. D{\bf 96}, 083520 (2017);  T. Zhu, A. Wang, K. Kirsten, G. Cleaver, and Q. Sheng, {\em Universal features of quantum bounce in loop quantum cosmology}, Phys. Lett.  B{\bf 773} (2017) 196. 
  
 \bibitem{tao2018}   T. Zhu, A. Wang,  K. Kirsten, G. Cleaver, and Q. Sheng, {\em Primoridial non-Gaussianity and power asymmetry with quantum gravitational effects in loop quantum cosmology}, Phys. Rev. D{\bf 97}, 043501 (2018); Q. Wu, T. Zhu, and A. Wang, {\em Nonadiabatic evolution of primordial perturbations and non-Gaussianity in hybrid approach of loop quantum cosmology}, Phys. Rev. D{\bf 98}, 103528 (2018).

\bibitem{abs2018}  I. Agullo, B. Bolliet and V. Sreenath,  {\em Non-Gaussianity in loop quantum cosmology}, Phys. Rev. D{\bf 97}, 066021 (2018). 
    

  \bibitem{nbm2018}  B.  Elizaga Navascu\'es, D. Mart\'in de Blas and G. A. Mena Marug\'an, {\em Time-dependent mass of cosmological perturbations in the hybrid and dressed metric approaches to loop quantum cosmology}, Phys. Rev. D{\bf 97}, 043523 (2018).
  
   \bibitem{nbm2018a} B.  Elizaga Navascu\'es, D. Mart\'in de Blas and G. A. Mena Marug\'an, {\em The vacuum state of primordial fluctuations in hybrid loop quantum cosmology}, Universe {\bf 4}, 98 (2018).
  
  \bibitem{lang94}  D. Langlois, {\em Hamiltonian formalism and gauge invariance for linear perturbations in inflation}, Class. Quant. Grav. {\bf11}, 389 (1994).

\bibitem{giesel1}  K.~Giesel, A.~Herzog and P.~Singh, {\em Gauge invariant variables for cosmological perturbation theory using geometrical clocks,}
  Class.\ Quant.\ Grav.\  {\bf 35}, 155012 (2018).

\bibitem{giesel2} K.~Giesel, P.~Singh and D.~Winnekens, {\em 
  Dynamics of Dirac observables in canonical cosmological perturbation theory,}
  Class.\ Quant.\ Grav.\  {\bf 36}, 085009 (2019).

\bibitem{b1-class}  I. Agullo, J. Olmedo and V. Sreenath, {\em Hamiltonian theory of classical and quantum gauge invariant perturbations in Bianchi I spacetimes}, Phys. Rev. D{\bf 101}, 123531 (2020).
  

      
    \bibitem{lsw2020} B. F. Li, P.~Singh, A. Wang, {\em Primordial power spectrum from the dressed metric approach in loop cosmologies}, Phys. Rev. D{\bf 100}, 086004 (2020).
    
    \bibitem{IA19}  I. Agullo, {\em Primordial power spectrum from the Dapor-Liegener model of loop quantum cosmology}, 
 Gen. Rel. Grav. {\bf 50}  (2018) 91.
  
  \bibitem{hawking} J. J. Halliwell and S. W. Hawking, {\em Origin of structure in the universe}, Phys. Rev. D{\bf 31}, 1777 (1985).
  
  \bibitem{mmo2012}  M. Fern\'andez-M\'endez,  G. A. Mena Marug\'an and  J. Olmedo  {\em Unique Fock quantization of scalar cosmological  perturbations}, Phys. Rev. D{\bf 85}, 103525 (2012).
  
    \bibitem{qm2019}  A. Garc\'ia-Quismondo and  G. A. Mena Marug\'an, {\em Martin-Benito-Mena Marugan-Olmedo prescription for the Dapor-Liegener model of loop quantum cosmology},  Phys. Rev. D{\bf 99}, 083505 (2019).

\bibitem{gqm2020} L. Castell\'o Gomar, A. Garc\'ia-Quismondo and G. A. Mena Marug\'an, {\em Primordial perturbations in the Dapor-Liegener model of hybrid loop quantum cosmology}, arXiv: 2002.01262.

\bibitem{qmp2020} A. Garc\'ia-Quismondo,  G. A. Mena Marug\'an and G. S\'anchez P\'erez, {\em The time-dependent mass of cosmological perturbations in loop quantum cosmology: Dapor-Liegener regularization}, arXiv: 2006.09781.

\bibitem{mop2011}  G. A. Mena Marug\'an, J. Olmedo and T. Pawlowski,  {\em Prescription in loop quantum cosmology: A comparative analysis},  Phys. Rev. D{\bf 84}, 064012 (2011).

\bibitem{bmo2009}  M. Mart\'in-Benito, G. A. Mena Marug\'an and J. Olmedo, {\em Further improvement in the understanding of isotropic loop quantum cosmology},  Phys. Rev. D{\bf 80}, 104015 (2009).


    \bibitem{VT08} V. Taveras, {\em Corrections to the Friedmann equations from LQG for a Universe with a free scalar field},   Phys. Rev. D{\bf 78}, 064072 (2008).
    
          \bibitem{Planck2018} P.~A.~R.~Ade {\it et al.} [Planck Collaboration],
  {\em Planck 2018 results. X. Constraints on inflation,} arXiv: 1807.06211.
    

\end{thebibliography}
\end{document}